\documentstyle[12pt]{article}

\def\fnote#1{\footnote}

\font\tenbm=cmmib10
\font\sevenbm=cmmib7
\font\fivebm=cmmib5
\newfam\bmfam
\textfont\bmfam=\tenbm \scriptfont\bmfam=\sevenbm
\scriptscriptfont\bmfam=\fivebm
{\count0=\number\bmfam \multiply\count0 by "100
\def\defbgreek#1#2#3{{\count1=\count0 \advance\count1 by "#2#3
  \global\mathchardef#1=\count1 }}
\defbgreek\balpha  0B \defbgreek\brho       1A
\defbgreek\bbeta   0C \defbgreek\bsigma     1B
\defbgreek\bgamma  0D \defbgreek\btau       1C
\defbgreek\bdelta  0E \defbgreek\bupsilon   1D
\defbgreek\bepsilon0F \defbgreek\bphi       1E
\defbgreek\bzeta   10 \defbgreek\bchi       1F
\defbgreek\bmeta   11 \defbgreek\bpsi       20
\defbgreek\btheta  12 \defbgreek\bomega     21
\defbgreek\biota   13 \defbgreek\bvarepsilon22
\defbgreek\bkappa  14 \defbgreek\bvartheta  23
\defbgreek\blambda 15 \defbgreek\bvarpi     24
\defbgreek\bmu     16 \defbgreek\bvarrho    25
\defbgreek\bnu     17 \defbgreek\bvarsigma  26
\defbgreek\bxi     18 \defbgreek\bvarphi    27
\defbgreek\bpi     19}
\normalbaselines

\begin{document}

\title{Integration of Complete System of Dynamic Equations for Ideal Fluid}
\author{Yuri A.Rylov}
\date{Institute for Problems in Mechanics, Russian Academy of Sciences,\\
101, bild.1 Vernadskii Ave., Moscow, 117526, Russia.}
\maketitle

\begin{abstract}
The Eulerian system of dynamic equations for the ideal (nondissipative)
fluid is closed but incomplete. The complete system of dynamic equations
arises after appending Lin constraints which describe motion of fluid
particles in a given velocity field. The complete system of dynamic
equations fo the ideal fluid can be integrated. Description in terms of
hydrodynamic potentials (DTHP) arises as a result of this integration. The
integrated system contains indefinite functions of three arguments, which
can be expressed via initial and boundary conditions. The remaining initial
and boundary conditions for the integrated system can be made universal
(i.e. the same for all fluid flows), and the resulting system of equations
contains full information about the fluid flow including initial and
boundary conditions for the fluid flow. Some hydrodynamic potentials appear
to be frozen into the fluid, and the Kelvin's theorem on the velocity
circulation can be formulated in a contour-free form. Description in terms
of the wave function (DTWF) appears to be a kind of DTHP. Calculation of
slightly rotational flows can be carried out on the basis of DTHP, or DTWF.
Such a description of a rotational flow appears to be effective.
\end{abstract}

\section{Introduction}

Dynamic equations for an ideal (nondissipative) fluid are written
conventionally in the form:
$$
{\frac{\partial \rho }{\partial t}}+\nabla (\rho {\bf v})=0 \eqno (1.1)
$$
$$
{\frac{\partial {\bf v}}{\partial t}}+({\bf v\nabla }){\bf v}=-{\frac{1}{\rho%
}} \nabla p, \qquad p=\rho ^2{\frac{\partial E}{\partial\rho}} \eqno (1.2)
$$
$$
{\frac{\partial S}{\partial t}}+({\bf v\nabla })S=0 \eqno (1.3)
$$
\noindent where dependent variables $\rho $ and ${\bf v}=\{v^{1},v^{2},v^{3}%
\}$, $S$ are respectively the fluid mass density, the fluid velocity and
entropy per unit mass considered as functions of independent Eulerian
variables $x=\{ t,{\bf x}\}$. $p$ is athe pressure, and $E=E(\rho ,S)$ is an
internal energy of an unite mass considered as a function of $\rho $ and $S$%
. The internal energy $E=E(\rho ,S)$ is an unique characteristic of the
ideal fluid.

The system of hydrodynamic equations (1.1)--(1.3) is a closed system of
differential equations which has an unique solution inside some space-time
region $\Omega$, provided dependent dynamic variables $\rho $ and ${\bf v}%
=\{v^{1},v^{2},v^{3}\}$, $S$ are given as functions of three arguments on
the space-time boundary $\Gamma$ of the region $\Omega$. Being closed, the
system (1.1)--(1.3) describes nevertheless only momentum-energetic
characteristics of the fluid. Motion of the fluid particles along
trajectories is described by so called Lin (1963) constraints
$$
{\frac{\partial\bxi}{\partial t}}+({\bf v}\nabla )\bxi =0, \eqno (1.4)
$$
where quantities $\bxi =\bxi (t,{\bf x}) =\{\xi _\alpha (t,{\bf x})\}$, $%
\alpha =1,2,3$ label fluid particles. They will be referred to as particle
labeling (curvilinear Lagrangian coordinates). If the equations (1.4) are
solved and $\bxi$ is determined as a function of $(t,{\bf x})$, the finite
relations
$$
\bxi (t,{\bf x})=\bxi _{{\rm in}}=\hbox{const } \eqno (1.5)
$$
\noindent  describe implicitly a fluid particle trajectory and a motion
along it.

The system of eight equations (1.1) --(1.4) forms a complete system of
dynamic equations describing a fluid, whereas the system of five equations
(1.1)--(1.3) forms a curtailed system of dynamic equations. The last system
is closed, but to be a complete system, it must be supplemented by the
kinematic equations
$$
{\frac{d{\bf x}}{dt}}={\bf v}(t,{\bf x}),\qquad {\bf x}={\bf x}(t,\bxi ) %
\eqno (1.6)
$$
\noindent where ${\bf v}(t,{\bf x})$ is a solution of the system
(1.1)-(1.3). Three equations (1.6) are equivalent to (1.4), because any
solution $\bxi =\bxi (t,{\bf x})$ of (1.4) is a set of three integrals of
equations (1.6).

There is a lack of understanding of the fact that the Euler system
(1.1)--(1.3) is a curtailed one and that the equations (1.4) describe a
motion of fluid particles. Even Lin (1963) who introduced equations (1.4)
considered them ''as the condition for the conservation of the identity of
particles'', but not as kinematic equations describing a motion of fluid
particles in a given velocity field. Betherton (1970) 
investigated the Lin constraints and showed a connection between them and
the Kelvin's theorem on the velocity circulation. But the very important
statement that the Lin constraints (1.4) are a kind of kinematic equations
(1.6) was not mentioned directly. In general, we failed to find in
literature any reference to equations (1.4) as necessary kinematic
equations, although this fact seems to be evident. Apparently, it is
connected with the common belief that the Lin constraints (1.4) are useless,
if one is interested only in the velocity field of the flow.

We should like to show a necessity of these equations in a very simple
example, modelling a situation in the hydrodynamics in a grotesque form. Let
us consider a particle moving in uniform gravitational field ${\bf g}=$%
const. Dynamic equations have the form
$$
\dot{{\bf v}}={\bf g},\qquad \dot{{\bf x}}={\bf v}, \qquad {\bf g}={\rm const%
} \eqno (1.7)
$$
where ${\bf x}$ and ${\bf v}$ are functions of $t$ describing respectively
position and velocity of the particle. First three equations (1.7)
constitute a closed subsystem [analog of the Euler system (1.1)--(1.3)] of
the full system of six equations. Equations of the subsystem can be solved
independently of the remaining equations. But it does not mean that this
closed subsystem may be considered as a system of dynamic equations
describing a particle, even if we are interested only in momentum-energetic
properties of the particle. Such important characteristic of the particle as
the energy integral ${\bf v}^2/2-{\bf g}{\bf x} ={\rm const} $ cannot be
derived on the base of only first three equations (1.7). This integral is a
sum of integrals
$$
(v_\alpha )^2/2-g_\alpha x_\alpha =C_\alpha ={\rm const}, \qquad v_\alpha ={%
\frac{dx^\alpha}{dt}},\qquad\alpha =1,2,3 \eqno (1.8)
$$
These integrals are analogs of integrals of the complete system (1.1)-(1.4)
[see below (1.11)]. This example shows that a closed system of equations and
a complete system of dynamic equations is not the same. The Lagrangian
formulation of hydrodynamic equation includes equations (1.4) in the form
(1.6) automatically. Lagrangian formulation is equivalent to the Eulerian
formulation (1.1)--(1.3), provided equations (1.4) are appended to it.

Equations (1.1), (1.3) can be integrated on the basis of (1.4) in the form
$$
S(t,{\bf x})=S_0(\bxi ) \eqno (1.9)
$$
$$
\rho (t,{\bf x})=\rho _0(\bxi ){\frac{\partial (\xi _1,\xi _2,\xi _3)}{%
\partial (x^1,x^2,x^3)}}\equiv\rho _0(\bxi ){\frac{\partial (\bxi )}{%
\partial ({\bf x})}} \eqno (1.10)
$$
where $S_0(\bxi )$ and $\rho _0(\bxi )$ are arbitrary integration functions
of the argument $\bxi$. These functions can be determined from the initial
(and boundary) conditions. Three equations (1.2) also can be integrated on
the basis of (1.4). These integrals have the form
$$
{\bf v}(t, {\bf x})= {\bf u}(\varphi ,\bxi ,\eta ,S)\equiv \nabla\varphi
+g^{\alpha }(\bxi ) \nabla\xi _{\alpha }-\eta\nabla S, \eqno (1.11)
$$
where $g(\bxi )=\{g^\alpha (\bxi )\}$, $\alpha =1,2,3$ are arbitrary
integration functions of argument $\bxi$, and $\varphi$, $\eta$ are new
dependent variables satisfying dynamic equations
$$
{\frac{\partial\varphi}{\partial t}}+{\bf u}(\varphi ,\bxi ,\eta
,S)\nabla\varphi -{\frac{1}{2}}{[{\bf u}(\varphi ,\bxi ,\eta ,S)]}^2 +{\frac{%
\partial (\rho E)}{\partial \rho }}=0 \eqno (1.12)
$$
$$
{\frac{\partial\eta}{\partial t}}+ {\bf u}(\varphi ,\bxi ,\eta ,S)\nabla\eta
= -{\frac{\partial E}{\partial S}}. \eqno (1.13)
$$
If five dependent variables $\varphi$, $\bxi$, $\eta$ satisfy the system of
equations (1.4), (1.12), (1.13), the five dynamic variables $S$, $\rho$, $%
{\bf v}$ (1.9)--(1.11) satisfy dynamic equations (1.1)--(1.3). Indefinite
functions $g^\alpha (\bxi )$ can be determined from initial and boundary
conditions in such a way that the initial and boundary conditions for
variables $\varphi$, $\bxi$, $\eta$ were universal in the sense that they do
not depend on the fluid flow.

According to (1.10), (1.11) the physical quantities $\rho $, ${\bf v}$ are
obtained as a result of differentiation of the variables $\varphi $, $\bxi$,
$S$, and the variables $\varphi $, $\bxi$, $\eta $ can be regarded as
hydrodynamic potentials. These potentials appear in the Hamilton fluid
dynamics (Salmon, 1988) as dependent variables. The hydrodynamic potentials
arise by a natural way. They associate with the name of Clebsch (1857, 1859)
who introduced these quantities for the incompressible fluid. Such
quantities as $g^\alpha (\bxi )$ also appear in the Hamilton fluid mechanics
(Salmon,1988), 
but they appear as dependent variables (Lagrage invariants) satisfying
dynamic equations of the type (1.4). They also are regarded as hydrodynamic
potentials. Note that in the Hamilton fluid mechanics the quantities $%
g^\alpha $ are considered simply as dependent variables, but not as
indefinite functions of $\bxi$, arising as a result of integration, although
corresponding dynamic equations for $g^\alpha $ can be integrated easily.

It should distinguish between the integration and a change of variables
which does not contain arbitrary functions explicitly. For instance, let us
substitute $g^\alpha (\bxi )$, $\alpha =1,2,3$ by new dependent variables $%
A^\alpha$, $\alpha =1,2,3$, imposing on them constraints
$$
{\frac{\partial A^\alpha}{\partial t}}+{\bf u}(\varphi ,\bxi ,\eta ,S)
\nabla A^\alpha =0, \qquad \alpha =1,2,3. \eqno (1.14)
$$
Then one has instead of (1.11)
$$
{\bf v}(t, {\bf x})= {\bf u}(\varphi ,\bxi ,\eta ,S)\equiv \nabla\varphi
+A^{\alpha } \nabla\xi _{\alpha }-\eta\nabla S, \eqno (1.15)
$$

The eighth order system (1.4), (1.12)--(1.15) arises instead of the fifth
order system (1.4), (1.11)--(1.13). This system is not an integrated one,
because it has the higher order and contains only two arbitrary functions
(1.9), (1.10). Different modifications of such kind systems were derived
(Salmon, 1988). They cannot be considered as integrated systems. Of course,
one can easily integrate equtions (1.14) on the basis of (1.4), and return
to (1.11). But until this is not made and the number of dynamic variables is
not decreased, the system cannot be considered as integrated.

The integration of the complete system (1.1)-(1.4) and some corollaries of
this integration correlates with the Hamilton properties of the ideal fluid
[Harivel, (1955); Eckart, (1960); Seliger and Whithem, (1967), Salmon,
(1988); Zakharov and Kuznetsov (1997)]. It is connected with the fact that
the curtailed system (1.1)-(1.3) is not a Hamiltonian system in itself,
whereas the complete system (1.1)-(1.4) is a Hamiltonian one that can be
easily seen in the example (1.7). Constructing Hamiltonian mechanics of the
ideal fluid, one uses (implicitly or explicitly) the Lin constraints (or
part of them). It is this expansion of the curtailed system (but not
Hamiltonian properties) that is important for integration and derivation of
other useful results. To show this, the Hamiltonian technique and
Hamiltonian properties of the ideal fluid will not be used at all.

Note that the curtailed system (1.1)--(1.3) has the same order as the
integrated system (1.4), (1.12), (1.13), but it takes into account neither
initial conditions, nor kinematic equations (1.6). The fact that the ideal
fluid considered as a dynamic system admits both the curtailed system
(1.1)--(1.3) and the integrated system (1.4), (1.12), (1.13) is connected
closely with the group of the relabeling transformation (relabeling group)
$$
\xi _{\alpha }\to\tilde{\xi }_{\alpha }=\tilde{\xi }_{\alpha }(\bxi ),
\qquad D\equiv\det \parallel \partial \tilde{\xi }_{\alpha }/\partial \xi
_{\beta } \parallel\neq 0,\qquad \alpha ,\beta =1,2,3 \eqno (1.16)
$$
$$
\varphi =\xi _0\to\tilde{\xi }_0 =\tilde{\varphi }=\tilde{\xi }_0(\xi _0)
+a_0(\bxi ),\qquad \partial\tilde{\xi _0}/\partial\xi _0 > 0 \eqno (1.17)
$$
where $\xi =\{\xi _0,\bxi \}$ are curvilinear Lagrangian coordinates in the
space-time, $\tilde{\xi }=\{\tilde{\xi }_0,\tilde{\bxi }\}$ is another
system of curvilinear Lagrangian coordinates. $\tilde{\bxi }_0$ and $a_0$
are arbitrary functions of $\bxi$. $\tilde{\xi }_0$ is arbitrary function of
$\xi _0$. $\xi _0$ is a temporal coordinate, and $\bxi$ are spatial ones.

The relabeling group is a symmetry group of the fluid considered as a
dynamic system. This circumstance admits to integrate the complete system
(1.1)--(1.4). Any special particle labeling is unessential from physical
viewpoint. It is a reason why several equations (1.1)--(1.3) of the complete
system form a closed system describing conservation laws. The relabeling
group is used in hydrodynamics comparatively recently Eckart (1938, 1960),
Calkin (1963), Bretherton (1970), Friedman and Schutz (1978), Salmon (1982),
Zacharov and Kuznetsov (1997) and others.

The integrated system (1.4), (1.12), (1.13) looks more complicated, than the
curtailed system (1.1)--(1.3). It is quite natural, because the integrated
system is a complete system which contains five indefinite functions $\rho
_0 $, $S_0$, $g^\alpha $ describing initial and boundary conditions.

There is a common belief that the curtailed Eulerian system (1.1)--(1.3) is
sufficient for calculating fluid flows, and a use of the complete system
(1.1)--(1.4) is not necessary in most of cases. In general, there is a
puzzling question of such a kind. If both the Euler system (1.1)-(1.3) and
the integrated system (1.4), (1.12), (1.13) have the same order and the same
number of dependent variables, why should one consider the integrated system
(1.4), (1.12), (1.3) which looks more complicated, than the Euler system?
The answer is as follows. Although the Lin constraints describe mainly a
motion of fluid particles in the given velocity field, nevertheless the
integrated system contains additional information which is necessary for a
calculation of rotational fluid flows. This additional information concerns
fluid properties described by the Kelvin's theorem on the velocity
circulation. This theorem is an {\it attribute of the complete system}
(1.1)-(1.4), because it refers to the contour connected rigidly with the
fluid particles moving according to dynamic equations (1.6), or (1.4). In
general, the Kelvin's theorem cannot be formulated only in terms of the
velocity field. On the other hand, it is clear that the fluid properties
(vorticity frozen into fluid) described by the Kelvin's theorem are very
important for calculating rotational flows, whereas for irrotational flows
the constraints imposed by the Kelvin's theorem degenerate into identities
which are fulfilled automatically.

The conventional theory of fluid flows is based mainly on the Euler system
(1.1)-(1.3) which does not take into account constrains of the Kelvin's
theorem. Theory of irrotational flows has been developed well enough,
whereas that of rotational flows has been developed much slightly.
Apparently, it is connected with the fact that the Euler system (1.1)-(1.3)
does not contain enough information on the fluid properties, and proper
calculations of rotational flows are impossible in the scope of the
curtailed Euler system (1.1)-(1.3). Especially it concerns strongly
rotational (turbulent) flows. It will be shown in this paper that slightly
rotational flows of incompressible fluid can be calculated on the basis of
the integrated system (1.4), (1.12), (1.13) (DTHP or DTWF), where integrals
(1.11) are taken into account.

The integrals (1.11) can be interpreted as a contour-free form of the
Kelvin's theorem in application to the perfect fluid. Indeed, let us
multiply (1.11) by $d{\bf x}$ and integrate along a closed path ${\cal L}$.
One obtains
$$
C_{{\cal L}}\equiv \oint\limits^{}_{{\cal L}}{\bf v}d{\bf x}=
\oint\limits^{}_{{\cal L}_\bxi }g^\alpha (\bxi )d\xi _\alpha +
\oint\limits^{}_{{\cal L}_\bxi } \eta {\frac{\partial S_0(\bxi )}{%
\partial\xi _\alpha}} d\xi _\alpha \eqno (1.18)
$$
where ${\cal L}_\bxi $ is a mapping of the contour ${\cal L}$ in the ${\bf x}
$-space onto the $\bxi$-space of labels $\bxi$. This mapping
$$
{\cal L\leftrightarrow L}_\bxi .\qquad \bxi \rightarrow {\bf x}(t,\bxi %
),\qquad {\bf x\rightarrow \bxi }(t,{\bf x})\eqno(1.19)
$$
depends on time $t$. Let the contour ${\cal L}$ be coupled rigidly with
fluid particles and move with the fluid. It means that ${\cal L}_\bxi $ is
fixed, ${\bf x}(t,\bxi )$ satisfies (1.6), and the shape of ${\cal L}$
depends on time according to (1.19), (1.6). Let the flow be homoentropic ($S(%
\bxi )=$const identically), or the contour ${\cal L}_\bxi $ lies on the
surface $S(\bxi )=$const., the second integral in rhs of (1.18) vanishes,
and the circulation ${\cal \ C}_L$ does not depend on time.

Let now the infinitesimal contour ${\cal L}$ be a parallelogram made up by
two infinitesimal vectors $d{\bf x}_1$, $d{\bf x}_2$ and ${\cal L}_\bxi $ be
made up by two infinitesimal vectors $d{\bf \bxi }_1$, $d{\bf \bxi }_2$.
Then using Stokes's theorem, one derives from (1.18)
$$
\bomega d {\bf S}={\bf \Omega }(\bxi )d{\bf S}_\bxi ,\qquad \bomega =\nabla
\times {\bf v,\qquad \Omega }(\bxi )=\{\Omega _\alpha (\bxi )\},\qquad
\alpha =1,2,3 \eqno(1.20)
$$
$$
\Omega _\alpha =\varepsilon _{\alpha \beta \gamma }\Omega ^{\beta \gamma
}\qquad \alpha =1,2,3 \qquad d{\bf S}=d{\bf x}_1\times d{\bf x}_2,\qquad d%
{\bf S}_\bxi = d\bxi _1\times d\bxi _2
$$
$$
\Omega ^{\alpha \beta }=\frac{\partial g^\alpha (\bxi )}{\partial \xi _\beta
}-\frac{\partial g^\beta (\bxi )}{\partial \xi _\alpha },\qquad \alpha ,
\beta =1,2,3 \eqno(1.21)
$$
where $\varepsilon _{\alpha \beta \gamma }$ is the Levi-Chivita
pseudotensor, and a summation is produced (1-3) over repeated Greek indices.
d${\bf S}$ and d${\bf S}_\xi $ are infinitesimal area of contours ${\cal L}$
and ${\cal L}_\bxi $ respectively. The relation (1.20) is a local (or
contour-free) form of the Kelvin's theorem. The scalar $\bomega d{\bf S}$
conserves and does not depend on time, although $\bomega $ and $d{\bf S}$
individually depend on time. For any irrotational flow the relation (1.20)
degenerates into identity, and Kelvin's theorem may be ignored.

Note that ${\bf \Omega }(\bxi )$ may be regarded as a "frozen vorticity",
because it depends on time $t$ only via $\bxi$, and ${\bf \Omega}=\bomega$,
provided $\bxi ={\bf x}$. ${\bf \Omega}(\bxi )$ is a scalar in the ${\bf x}$%
-space, and it is a vector in the $\bxi$-space of labels $\bxi$.

For derivation of integrated system (1.4), (1.12), (1.13) one uses a
specific mathematical technique based on Jacobian properties. This technique
permits to integrate dynamic equations without a use of a change of
variables. In general, all results can be obtained, carrying out a proper
change of variables in the action functional by means of the Lagrange
multipliers. Such changes of dependent variables are produced in the
Hamilton fluid dynamics (see, for instance, Salmon, 1988). Dynamic equations
of the type of (1.3) appear as a result of such changes of variables. These
equations can be integrated easily on the basis of equations (1.4).
Unfortunately, such changes of variables lead to different sets of dependent
variables whose physical meaning is unclear. In other words, using Lagrange
multipliers, for a change of variables, one obscures logical connection
between different variables. It is rather difficult to understand that,
integrating some equations of the type of (1.3), one integrates in reality
the equations (1.2). To simplify the logical connection between different
dependent variables and to clear their physical meaning, we prefer to
integrate dynamic equations directly by means of ''Jacobian technique''
(sec. 2). Use of Jacobians in hydrodynamics has had a long history, dating
back to the time of Clebsch (1857, 1859). It was the use of Jacobians that
allowed to introduce the Clebsch potentials and integrate hydrodynamic
equations.

The Jacobian technique was used by many authors (Herivel (1955),
Eckart(1960), Berdichevski (1983), Salmon (1988), Zacharov and Kuznetsov
(1997) and many others). We use space-time symmetric version of the Jacobian
technique which appears to be simple and effective. It seems that the
progress in the integration of hydrodynamic equations is connected mainly
with the developed Jacobian technique.

\section{Jacobian technique}

Let us consider such a space-time symmetric mathematical object as the
Jacobian
$$
J\equiv {\frac{\partial (\xi _0,\xi _1,\xi _2,\xi _3)}{\partial
(x^0,x^1,x^2,x^3)}}\equiv \det \parallel \xi _{i,k}\parallel ,\qquad \xi
_{i,k}\equiv \partial _k\xi _i\equiv {\frac{\partial \xi _i}{\partial x^k}}%
,\qquad i,k=0,1,2,3\eqno (2.1)
$$
Here $\xi =\{\xi _0,\bxi\}=\{\xi _0,\xi _1,\xi _2,\xi _3\}$ are four scalar
considered as functions of $x=\{x^0,{\bf x}\}$, $\xi =\xi (x)$. The
functions $\{\xi _0,\xi _1,\xi _2,\xi _3\}$ are supposed to be independent
in the sense that $J\neq 0$. It is useful to consider the Jacobian $J$ as
4-linear function of variables $\xi _{i,k}\equiv \partial _k\xi _i$, $%
i,k=0,1,2,3$. Then one can introduce derivatives of $J$ with respect to $\xi
_{i,k}$. The derivative $\partial J/\partial \xi _{i,k}$ appears as a result
of a substitution of $\xi _i$ by $x^k$ in the relation (2.1).

$$
{\frac{\partial J}{\partial \xi _{i,k}}}\equiv {\frac{\partial (\xi
_0,...\xi _{i-1},x^k,{\xi _{i+1},...}\xi _3)}{\partial (x^0,x^1,x^2,x^3)}}%
,\qquad i,k=0,1,2,3\eqno (2.2)
$$
For instance
$$
{\frac{\partial J}{\partial \xi _{0,i}}}\equiv {\frac{\partial (x^i,\xi
_1,\xi _2,\xi _3)}{\partial (x^0,x^1,x^2,x^3)}},\qquad i=0,1,2,3\eqno (2.3)
$$
This rule is valid for higher derivatives of $J$ also.
$$
{\frac{\partial ^2J}{\partial \xi _{i,k}{\partial \xi _{s,l}}}}\equiv {\frac{%
\partial (\xi _0,...\xi _{i-1},x^k,{\xi _{i+1},...\xi _{s-1},x^l,{\xi
_{s+1},...}}\xi _3)}{\partial (x^0,x^1,x^2,x^3)}}\equiv
$$
$$
{\frac{\partial (x^k,x^l)}{\partial (\xi _i,\xi _s)}} \frac{\partial (\xi
_0,\xi _1,\xi _2,\xi _3)}{\partial (x^0,x^1,x^2,x^3)}\equiv J({\frac{%
\partial x^k}{\partial \xi _i}\frac{\partial x^l}{\partial \xi _s}- \frac{%
\partial x^k}{\partial \xi _s}\frac{\partial x^l}{\partial \xi _i})},\qquad
i,k,l,s=0,1,2,3\eqno (2.4)
$$
It follows from (2.1), (2.2) that
$$
{\frac{\partial x^k}{\partial \xi _i}}\equiv {\frac{\partial (\xi _0,...\xi
_{i-1},x^k,{\xi _{i+1},...}\xi _3)}{\partial {(\xi _0,\xi _1,\xi _2,\xi _3)}}%
}\equiv {\frac{\partial (\xi _0,...\xi _{i-1},x^k,{\xi _{i+1},...}\xi _3)}{%
\partial (x^0,x^1,x^2,x^3)}}\times
$$
$$
\frac{{\partial (x^0,x^1,x^2,x^3)}}{{\partial {(\xi _0,\xi _1,\xi _2,\xi _3)}%
}}\equiv {\frac 1J\frac{\partial J}{\partial \xi _{i,k}}},\qquad i,k=0,1,2,3%
\eqno (2.5)
$$
and (2.4) may be written in the form
$$
{\frac{\partial ^2J}{\partial \xi _{i,k}\partial \xi _{s,l}}}\equiv {\frac 1J%
}({\frac{\partial J}{\partial \xi _{i,k}}}{\frac{\partial J}{\partial \xi
_{s,l}}}-{\frac{\partial J}{\partial \xi _{i,l}}}{\frac{\partial J}{\partial
\xi _{s,k}}}),\qquad i,k,l,s=0,1,2,3\eqno (2.6)
$$

The derivative $\partial J/\partial \xi _{i,k}$ is a cofactor to the element
$\xi _{i,k}$ of the determinant (2.1). Then one has the following identities
$$
\xi _{l,k}{\frac{\partial J}{\partial \xi _{s,k}}}\equiv \delta _l^sJ,\qquad
\xi _{k,l}{\frac{\partial J}{\partial \xi _{k,s}}}\equiv \delta _l^sJ,\qquad
l,s=0,1,2,3\eqno (2.7)
$$
$$
\partial _k{\frac{\partial J}{\partial \xi _{i,k}}}\equiv {\frac{\partial
^2J }{\partial \xi _{i,k}\partial \xi _{s,l}}}\partial _k\partial _l\xi
_s\equiv 0,\qquad i=0,1,2,3.\eqno (2.8)
$$
Here and further a summation on two repeated indices is produced (0-3) for
Latin indices and (1-3) for the Greek ones. The identity (2.8) can be
considered as a corollary of the identity (2.6) and a symmetry of $\partial
_k\partial _l\xi _s$ with respect to permutation of indices $k$, $l$.
Convolution of (2.6) with $\partial _k$, or $\partial _l$ vanishes also.

Relations (2.1) --(2.6) are written for four independent variables $x$, but
they are valid in an evident way for arbitrary number $n+1$ of variables $%
x=\{x^0,x^1,\ldots x^n\}$ and $\bxi =\{\xi _0,\bxi \}$, \quad $\bxi =\{\xi
_1,\xi _2,\ldots \xi _n\}$.

Application of the Jacobian $J$ to hydrodynamics is founded on the property,
that the fluid flux
$$
j^i=m{\frac{\partial J}{\partial \xi _{0,i}}},\qquad j=\{j^i\}=\{\rho ,\rho
{\bf v\}},\qquad i=0,1,2,3\eqno (2.9)
$$
\noindent constructed on the basis of the variables $\bxi =\{\xi _1,\xi
_2,\xi _3\}$ satisfies Lin constraints (1.4) and the continuity equation
$$
\partial _ij^i=0\eqno (2.10)
$$
\noindent identically for any choice of variables $\bxi $, as it follows
from the identity (2.8) for $i=0$. The continuity equation (2.10) is used
without approximations in all hydrodynamic models, and the change of
variables $\{\rho ,\rho {\bf v\}\leftrightarrow \bxi }$ described by (2.9)
is very important.

In particular, in the case of two-dimensional established flow of
incompressible fluid the variables $\bxi $ reduce to one variable $\xi
_{1}=\psi $, known as the stream function. In this case there are only two
essential dependent variables $x^{0}=x$, $x^{1}=y$, and the relations (2.9),
(2.10) reduce to relations
$$
\rho ^{-1}j_{x}=v_{x}={\frac{\partial \psi }{\partial y}}, \qquad \rho
^{-1}j_{y}=v_{y}=-{\frac{\partial \psi }{\partial x}},\qquad {\frac{\partial
v_{x}}{\partial x}}+{\frac{\partial v_{y}}{\partial y}}=0 \eqno (2.11)
$$
Defining the stream line as a line tangent to the flux $j$
$$
{\frac{dx}{j_{x}}}={\frac{dy}{j_{y}}}, \eqno (2.12)
$$
\noindent one obtains that the stream function is constant along the stream
line, because according to two first equations (2.11), $\psi =\psi (x,y)$ is
an integral of the equation (2.12).

In the general case, when the space dimensionality is $n$ and $%
x=\{x^{0},x^{1},\ldots x^{n}\}$, $\bxi =\{\xi _{0},\bxi \}$, $\bxi =\{\xi
_{1},\xi _{2},\ldots \xi _{n}\}$, the quantities $\bxi =\{\xi _{\alpha }\}$,
$\alpha =1,2,\ldots n$ are constant along the line ${\cal L}$ tangent to the
flux vector $j=\{j^i\}$, $i=0,1,\ldots n$
$$
{\cal L}:\qquad {\frac{dx^i}{d\tau}}=j^{i}(x),\qquad i=0,1,\ldots n \eqno %
(2.14)
$$
\noindent where $\tau$ is a parameter along the line ${\cal L}$ which is
described parametrically by the equation $x=x(\tau)$. This statement is
formulated mathematically in the form
$$
{\frac{d\xi _{\alpha }}{d\tau}}=j^{i}\partial _{i}\xi _{\alpha }= m{\frac{%
\partial J}{\partial \xi _{0,i}}}\partial _{i}\xi _{\alpha }=0, \qquad
\alpha =1,2,\ldots n \eqno (2.15)
$$
The last equality follows from the first identity (2.7) taken for $s=0$, $%
l=1,2,\ldots n$

Interpretation of the line (2.14) tangent to the flux is different for
different cases. If $x=\{x^0,x^1,\ldots x^n\}$ contains only spatial
coordinates, the line (2.14) is a line in the usual space. It is regarded as
a stream line, and $\bxi $ can be interpreted as quantities which are
constant along the stream line (i.e. as a generalized stream function). If $%
x^0$ is the time coordinate, the equation (2.14) describes a line in the
space-time. This line (known as a world line of a fluid particle) determines
a motion of the fluid particle. Variables $\bxi =\{\xi _1,\xi _2,\ldots \xi
_n\}$ which are constant along the world line are different, generally, for
different particles. If $\xi _\alpha $, $\alpha =1,2,\ldots n$ are
independent, they may be used for the fluid particle labeling.

Thus, although interpretation of the relation (2.9) considered as a change
of dependent variables $j$ by $\bxi $ may be different, from the
mathematical viewpoint this transformation means a replacement of the
continuity equation by some equations for the labeling (or generalized
stream function) $\bxi $. Difference of the interpretation is of no
importance in this context.

Note that the expressions
$$
j^{i}= m\rho _0(\bxi ){\frac{\partial J}{\partial \xi _{0,i}}} \equiv m\rho
_0(\bxi ){\frac{\partial (x^{i},\xi _{1},\xi _{2},\xi _{3})}{\partial
(x^{0},x^{1},x^{2},x^{3})}}, \qquad i=0,1,2,3, \eqno (2.16)
$$
can be also considered as four-flux satisfying the continuity equation
(2.10). Here $m$ is a constant and $\rho _0(\bxi )$ is an arbitrary function
of $\bxi$. It follows from the identity
$$
m\rho _0(\bxi ){\frac{\partial (x^{i},\xi _{1},\xi _{2},\xi _{3})}{\partial
(x^{0},x^{1},x^{2},x^{3})}}\equiv m{\frac{\partial (x^{i},\tilde{\xi }%
_{1},\xi _{2},\xi _{3})}{\partial (x^{0},x^{1},x^{2},x^{3})}}, \qquad \tilde{%
\xi }_1=\int\limits^{\xi_1}_0 \rho _0(\xi _1^\prime ,\xi _2,\xi _3) d\xi
_1^\prime . \eqno (2.17)
$$

As an example of application of the Jacobian technique, let us show that
(1.10) satisfies (1.1) in virtue of (1.4). Let us multiply (1.4) by (1.10)
and introduce new variables ${\bf j}=\rho{\bf v}=\{j^1,j^2,j^3\}$. One
obtains three equations
$$
m\rho _0(\bxi ){\frac{\partial J}{\partial \xi _{0,0}}}\xi _{\beta ,0}+
j^\alpha \xi _{\beta ,\alpha }=0, \qquad \beta =1,2,3. \eqno (2.18)
$$
Considering (2.18) as a system of three linear equations for $j^\alpha$, $%
\alpha =1,2,3$ and resolving it with respect to $j^\alpha$, one obtains
$$
j^{\alpha }= m\rho _0(\bxi ){\frac{\partial J}{\partial \xi _{0,\alpha }}},
\qquad \alpha =1,2,3 \eqno (2.19)
$$
It is easy to verify this, substituting (2.19) into (2.18) and using (2.7).
One obtains that $j=\{j^0 ,{\bf j}\}=\{\rho ,\rho {\bf v}\}$ is described by
the relations (2.16) which satisfy the continuity equation (2.10)
identically. Thus, (1.1) is satisfied by (1.10) in virtue of (1.4).

\section{Variational principle}

In general, equivalency of the system (1.4), (1.12), (1.13) and the system
(1.1)--(1.4) can be verified by a direct substitution of variables $\rho $, $%
S$, ${\bf v}$, defined by the relations (1.9)--(1.11), into the equations
(1.1)--(1.3). Using equations (1.4), (1.12), (1.13), one obtains identities
after subsequent calculations. But such computations do not display a
connection between the integration and the invariancy with respect to the
relabeling group (1.16). Besides a meaning of new variables $\varphi $, $%
\eta $ is not clear. We shall use for our investigations a variational
principle. Note that for a long time a derivation of a variational principle
for hydrodynamic equaitons (1.1)--(1.3) was existing as a self-dependent
problem (Davydov, 1949; 
Herivel, 1955; 
Eckart, 1960; 
Lin, 1963; 
Seliger and Whithem, 1967; 
Bretherton, 1970; 
Salmon, 1988). 
Existence of this problem was connected with a lack of understanding that
the system of hydrodynamic equations (1.1)--(1.3) is a curtailed system, and
the full system of dynamic equations (1.1)--(1.4) includes equations (1.4)
describing a motion of the fluid particles in the given velocity field. The
variational principle can generate only the complete system of dynamic
variables (but not its closed subsystem). Without understanding this one
tried to form the Lagrangian for the system (1.1)--(1.3) as a sum of some
quantities taken with Lagrange multipliers. lhs of dynamic equations
(1.1)--(1.3) and some other constraints were taken as such quantities.

Now this problem has been solved (see review by Salmon, 1988) on the basis
of the Eulerian version of the variational principle for the Lagrangian
description, where equations (1.4) appear automatically and cannot be
ignored. In our version of the variational principle we follow Salmon (1988)
with some modifications which underline a curtailed character of
hydrodynamic equations (1.1)--(1.3), because the understanding of the
curtailed character of the system (1.1)--(1.3) removes the problem of
derivation of the variational principle for the hydrodynamic equations
(1.1)--(1.3).

We consider the ideal fluid as a conservative dynamic system whose dynamic
equations can be derived from the variational principle. This dynamic system
is a continuous set of many identical particles moving in some
self-consistent (and external) potential force field. The action functional
has the form
$$
{\cal A}_{{\rm L}}[{\bf x} ]=\int \{{\frac{m}{2}}{({\frac{d{\bf x}}{dt}})}^2
-V\}\rho _0(\bxi )dtd\bxi, \eqno (3.1)
$$
where ${\bf x}=\{x^\alpha (t,\bxi )\}$, $\alpha =1,2,3$ are dependent
variables considered as functions of time $t$ and of labels (Lagrangian
coordinates) $\bxi =\{\xi _1,\xi _2,\xi _3\}$. $d{\bf x}/dt$ is a derivative
of ${\bf x}$ with respect to $t$ taken at fixed $\bxi$.
$$
{\frac{dx^\alpha }{dt}} \equiv {\frac{\partial (x^\alpha ,\xi _1,\xi _2,\xi
_3)}{\partial (t,\xi _1,\xi _2,\xi _3)}} \equiv {\frac{\partial (x^\alpha ,%
\bxi )}{\partial (t,\bxi )}}\qquad \alpha =1,2,3 \eqno (3.2)
$$
$\rho _0(\bxi )$ is some non-negative weight function, and $V$ is a
potential of a self-consistent force field which depends on $\bxi $, ${\bf x}
$ and derivatives of ${\bf x}$ with respect to $\bxi$. $m=$const is some
mass of the fluid particle. The form of the potential $V$ will be fixed
later. Now it is important only that $V$ does not depend on the time
derivatives of ${\bf x}$.

Variation of the action with respect to ${\bf x}$ generates six first order
dynamic equations for six dependent variables ${\bf x}$, ${\bf v}=d{\bf x}%
/dt $, considered as functions of $t$ and of independent curvilinear
Lagrangian coordinates $\bxi$. It is a Lagrangian representation of
hydrodynamic equations.

We prefer to work with Eulerian representation, when Lagrangian coordinates
(particle labeling) $\xi =\{\xi _0,\bxi \}$, $\bxi =\{\xi _1,\xi _2,\xi _3\}$
are considered as dependent variables, and Eulerian coordinates $x=\{x^0,%
{\bf x}\}=\{t,{\bf x}\}$, ${\bf x}=\{x^1,x^2,x^3\}$ are considered as
independent variables. Here $\xi _0$ is a temporal Lagrangian coordinate
which evolves along the particle trajectory in an arbitrary way. Now the $%
\xi _0$ is a fictive variable, but after integration of equations the $\xi
_0 $ stops to be fictive and turns to the variable $\varphi$, appearing in
the integrated system (1.4), (1.12), (1.13).

Further mainly space-time symmetric designations will be used, that
simplifies considerably all computations. In the Eulerian description the
action functional (3.1) is to be represented as an integral over independent
variables $x=\{x^0,{\bf x}\}=\{t,{\bf x}\}$. One uses the Jacobian technique
for such a transformation of the action (3.1),

Let us note that according to (2.3) the derivative (3.2) can be written in
the form
$$
v^\alpha ={\frac{dx^\alpha}{dt}}\equiv {\frac{\partial J}{\partial\xi
_{0,\alpha }}} \left( {\frac{\partial J}{\partial\xi _{0,0}}}\right) ^{-1},
\qquad \alpha =1,2,3. \eqno (3.3)
$$
Then components of the 4-flux $j=\{j^0,{\bf j}\}\equiv \{\rho ,\rho {\bf v}%
\} $ can be written in the form (2.16), provided the designation (1.10)
$$
j^{0}=\rho = m\rho _0(\bxi ){\frac{\partial J}{\partial \xi _{0,0}}} \equiv
m\rho _0(\bxi ){\frac{\partial (x^{0},\xi _{1},\xi _{2},\xi _{3})}{\partial
(x^{0},x^{1},x^{2},x^{3})}} \eqno (3.4)
$$
is used.

At such form of the mass density $\rho$ the four-flux $j=\{ j^i\}$, $%
i=0,1,2,3$ satisfies identically the continuity equation (2.10) which takes
place in virtue of identities (2.7), (2.8). Besides in virtue of identities
(2.7), (2.8) the Lin constraints (1.4) are fulfilled identically
$$
j^i\partial _i\xi _\alpha =0, \qquad \alpha =1,2,3. \eqno (3.5)
$$
Components $j^i$ are invariant with respect to the relabeling group (1.16),
provided the function $\rho _0(\bxi )$ transforms as follows
$$
\rho _0(\bxi )\to\tilde{\rho }_0(\tilde{\bxi })=D^{-1}\rho _0(\bxi ), \qquad
D={\frac{\partial (\tilde{\bxi })}{\partial (\bxi )}} \equiv {\frac{\partial
(\tilde{\xi _1},\tilde{\xi _2},\tilde{\xi _3})}{\partial (\xi _1,\xi _2,\xi
_3)}} \eqno (3.6)
$$

One has
$$
\rho _0(\bxi )dtd\bxi = \rho _0(\bxi ){\frac{\partial J}{\partial\xi _{0,0}}}
dtd{\bf x}={\frac{\rho}{m}}dtd{\bf x} \eqno (3.7)
$$
$$
{\frac{m}{2}}\left( {\frac{dx^\alpha}{dt}}\right) ^2 ={\frac{m}{2}}\left( {%
\frac{\partial J}{\partial \xi _{0,\alpha }}}\right)^2\left( {\frac{\partial
J}{\partial\xi _{0,0}}}\right) ^{-2}= {\frac{m}{2}}{\left( {\frac{j^\alpha}{%
\rho }}\right) }^2, \eqno (3.8)
$$
and the variational problem with the action functional (3.1) is written as a
variational problem with the action functional
$$
{\cal A}_{{\rm E}}[\bxi ]=\int ({\frac{{\bf j}^{2}}{2\rho }}-\rho E)dtd{\bf x%
}, \qquad E={\frac{V}{m}} \eqno (3.9)
$$
where $\rho =j^0$ and ${\bf j}=\{j^1,j^2,j^3\}$ are fixed functions of $\xi
=\{\xi _0,\bxi \}$ and of $\xi _{\alpha ,i}\equiv\partial _i\xi _\alpha$, $%
\alpha =1,2,3$, $i=0,1,2,3$, defined by the relations (2.16). $E$ is the
internal energy of the fluid which is supposed to be a fixed function of $%
\rho$ and $S_0(\bxi )$
$$
E=E(\rho ,S_0(\bxi )), \eqno (3.10)
$$
where $\rho $ is defined by (3.4) and $S_0(\bxi )$ is some fixed function of
$\bxi$, describing initial distribution of the entropy over the fluid.

The action (3.9) is invariant with respect to subgroup ${\cal G}_{S_0}$ of
the relabeling group (1.16). The subgroup ${\cal G}_{S_0}$ is determined in
such a way that any surface $S_0(\bxi )=$const is invariant with respect to $%
{\cal G}_{S_0}$. In general, the subgroup ${\cal G}_{S_0}$ is determined by
two arbitrary functions of $\bxi$.

The action (3.9) generates the six order system of dynamic equations,
consisting of three second order equations for three dependent variables $%
\bxi$. Invariancy of the action (3.9) with respect to the subgroup ${\cal G}%
_{S_0}$ admits one to integrate the system of dynamic equations. The order
of the system is reduced, and two arbitrary integration functions appear.
The order of the system is reduced to five (but not to four), because the
fictive dependent variable $\xi _0$ stops to be fictive as a result of the
integration.

Unfortunately, the subgroup ${\cal G}_{S_0}$ depends on the form of the
function $S_0(\bxi )$ and cannot be obtained in a general form. In the
special case, when $S_0(\bxi )$ does not depend on $\bxi$, the subgroup $%
{\cal G}_{S_0}$ coincides with the whole relabeling group ${\cal G}$, and
the order of the integrated system is reduced to four.

In the general case it is convenient to introduce a new dependent variable
$$
S=S_0(\bxi ). \eqno (3.11)
$$
According to (3.5) the variable $S$ satisfies the dynamic equation (1.3)
$$
j^i\partial _iS=0. \eqno (3.12)
$$
In virtue of designations (2.9) and identities (2.7), (2.8) the equations
(3.12), (3.5) are fulfilled identically. Hence, they can be added to the
action functional (3.9) as side constraints without a change of the
variational problem. Adding (3.12) to the Lagrangian of the action (3.9) by
means of a Lagrange multiplier $\eta$, one obtains
$$
{\cal A}_{{\rm E}}[\bxi ,\eta , S]=\int \{ {\frac{{\bf j}^{2} }{2\rho }}%
-\rho E+\eta j^k\partial _kS\}dtd{\bf x} \eqno (3.13)
$$
where the quantities $j=\{\rho ,{\bf j}\}$ are determined by (2.16), and $%
E=E(\rho ,S)$. The action (3.13) is invariant with respect to the relabeling
group ${\cal G}$ which is determined by three arbitrary functions of $\bxi$.
Three arbitrary functions of $\bxi$ appear in consequence of the integration
of dynamic equations. The integrated system contains five first order
equations for dependent variables $\xi _0$, $\bxi$, $\eta$. The dependent
variable $S$ is substituted by arbitrary indefinite function $S_0(\bxi )$.

To obtain the dynamic equations, it is convenient to introduce new dependent
variables $j^i$, defined by (2.16). Let us introduce the new variables $j^i$
by means of designations (2.16) taken with the Lagrange multipliers $p_i$, $%
i=0,1,2,3$. Then the action (3.13) takes the form
$$
{\cal A}_{{\rm E}}[\rho ,{\bf j},\bxi ,p,\eta , S]=\int \{ {\frac{{\bf j}%
^{2} }{2\rho }}- \rho E-p_{k}[j^{k}-m \rho _0(\bxi ){\frac{\partial J}{%
\partial \xi _{0,k}}}] +\eta j^k\partial _kS\}dtd{\bf x} \eqno (3.14)
$$
It is useful to keep in mind that four designations (2.16), introducing
variables $\rho$, ${\bf j}=\rho {\bf v}$ via variables $\bxi$, are
equivalent to three Lin constraints (1.4) together with the designation
(3.4), as it was shown in the end of sec.2. Addition of relations (2.16) to
the action (3.13) as side constraints is equivalent to the addition of
relations (1.4), (3.4) considered as side constraints.

For obtaining dynamic equations the variables $\rho ,{\bf j},\bxi ,p,\eta ,S$
are to be varied. Let us eliminate the variables $p_i$ from the action
(3.14). Dynamic equations arising as a result of a variation with respect to
$\xi _\alpha $ have the form
$$
{\frac{\delta {\cal A}_{{\rm E}}}{\delta \xi _\alpha }}\equiv \hat {{\cal L}%
}_\alpha p=-m\partial _k[\rho _0(\bxi ){\frac{\partial ^2J}{\partial \xi
_{0,i}\partial \xi _{\alpha ,k}}}p_i]+m{\frac{\partial \rho _0(\bxi )}{%
\partial \xi _\alpha }}{\frac{\partial J}{\partial \xi _{0,k}}}p_k=0,\qquad
\alpha =1,2,3\eqno (3.15)
$$
where $\hat {{\cal L}}_\alpha $ are linear operators acting on variables $%
p=\{p_i\}$, $i=0,1,2,3$. This equations can be integrated in the form
$$
p_i=g^0(\xi _0)\partial _i\xi _0+g^\alpha (\bxi )\partial _i\xi _\alpha
,\qquad i=0,1,2,3,\eqno (3.16)
$$
\noindent where $\xi _0$ is some new variable (temporal Lagrangian
coordinate), $g^\alpha (\bxi )$, $\alpha =1,2,3$ are arbitrary functions of
the label $\bxi $, $g^0(\xi _0)$ is an arbitrary function of $\xi _0$. The
relations (3.16) satisfy equations (3.15) identically. Indeed, substituting
(3.16) into (3.15) and using identities (2.6), (2.7), one obtains
$$
-m\partial _k\left\{ \rho _0(\bxi )\left[ {\frac{\partial J}{\partial \xi
_{\alpha ,k}}}g^0(\xi _0)-{\frac{\partial J}{\partial \xi _{0,k}}}g^\alpha (%
\bxi )\right] \right\} +m{\frac{\partial \rho _0(\bxi )}{\partial \xi
_\alpha }}Jg^0(\xi _0)=0,\qquad \alpha =1,2,3,\eqno (3.17)
$$
Differentiating braces and using identities (2.8), (2.7), one concludes that
(3.17) is an identity.

Setting for simplicity
$$
\partial _{k}\varphi =g^{0}(\xi _{0})\partial _{k}\xi _0,\qquad k=0,1,2,3 %
\eqno (3.18)
$$
\noindent one obtains
$$
p_{k}=\partial _{k}\varphi +g^{\alpha }(\bxi )\partial _{k}\xi _{\alpha },
\qquad k=0,1,2,3 \eqno (3.19)
$$

Substituting (3.19) in (3.14), one can eliminate variables $p_i$, $i=0,1,2,3$
from the functional (3.14). The term $g^\alpha (\bxi )\partial _k\xi _
\alpha {\partial J/\partial \xi _{0,k}}$ vanish, the term $\partial
_k\varphi {\partial J/\partial \xi _{0,k}}$ gives no contribution into
dynamic equations. The action functional takes the form
$$
{\cal A}_{{\bf g}}[\rho ,{\bf j},\bxi ,\eta ,S]=\int \{{\frac{{\bf j}^2}{%
2\rho }}-\rho E-j^k[\partial _k\varphi +g^\alpha (\bxi )\partial _k\xi
_\alpha -\eta \partial _kS]\}dtd{\bf x}\eqno (3.20)
$$
where $g^\alpha (\bxi )$ are considered as fixed functions of $\bxi$ which
are determined from initial conditions. Varying the action (3.20) with
respect to $\varphi $, $\bxi$, $\eta $, $S$, ${\bf j}$, $\rho $, one obtains
dynamic equations
$$
\delta \varphi :\qquad \partial _kj^k=0,\eqno (3.21)
$$
$$
\delta \xi _\alpha :\qquad \Omega ^{\alpha \beta }j^k\partial _k\xi _\beta
=0,\qquad \alpha =1,2,3,\eqno (3.22)
$$
where $\Omega ^{\alpha\beta }$ is defined by (1.21)
$$
\delta \eta :\qquad j^k\partial _kS=0,\eqno (3.23)
$$
$$
\delta S:\qquad j^k\partial _k\eta =-\rho {\frac{\partial E}{\partial S}},%
\eqno (3.24)
$$
$$
\delta {\bf j}:\qquad {\bf v}\equiv {\bf j}/\rho =\nabla \varphi +g^\alpha (%
\bxi )\nabla \xi _\alpha -\eta \nabla S,\eqno (3.25)
$$
$$
\delta \rho :\qquad -{\frac{{\bf j}^2}{2\rho ^2}}-{\frac{\partial (\rho E)}{%
\partial \rho }}-\partial _0\varphi -g^\alpha (\bxi )\partial _0\xi _\alpha
+\eta \partial _0S=0,\eqno (3.26)
$$
Deriving relations (3.22), (3.24), the continuity equation (3.21) was used.
It is easy to see that (3.22) is equivalent to (1.4), provided
$$
\det \parallel \Omega ^{\alpha \beta }\parallel \neq 0\eqno (3.27)
$$

Then the equations (3.23) and (3.21) can be integrated in the form of (1.9)
and (1.10) respectively. Equations (3.24) and (3.25) are equivalent to
(1.13) and (1.11). Finally, eliminating $\partial _0\xi _\alpha$ and $%
\partial _0S$ from (3.26) by means of (3.22) and (3.23), one obtains the
equation (1.12) and, hence, the system of dynamic equations (1.4), (1.12),
(1.13), where designations (1.9)--(1.11) are used.

The curtailed system (1.1)--(1.3) can be obtained from equations
(3.21)--(3.26) as follows. Equations (3.21), (3.23) coincide with (1.1),
(1.3). For deriving (1.2) let us note that the vorticity $\bomega %
\equiv\nabla\times {\bf v} $ and ${\bf v}\times\bomega  $ are obtained from
(3.25) in the form
$$
\bomega =\nabla\times {\bf v}={\frac{1}{2}} {\Omega }^{\alpha\beta
}\nabla\xi _\beta \times\nabla\xi _\alpha -\nabla\eta\times\nabla S \eqno %
(3.28)
$$
$$
{\bf v}\times\bomega = {\Omega }^{\alpha\beta }\nabla \xi _\beta ({\bf v}%
\nabla )\xi _\alpha +\nabla S({\bf v}\nabla )\eta -\nabla\eta ({\bf v}\nabla
)S \eqno (3.29)
$$

Let us form a difference between the time derivative of (3.25) and the
gradient of (3.26). Eliminating ${\Omega }^{\alpha\beta }\partial _0\bxi %
_\alpha$, $\partial _0S$ and $\partial _0\eta$ by means of equations (3.22),
(3.23), (3.24), one obtains
$$
\partial _0{\bf v}+\nabla {\frac{{\bf v}^2}{2}}+ {\frac{\partial ^2(\rho E)}{%
\partial\rho ^2 }}\nabla\rho + {\frac{\partial ^2(\rho E)}{%
\partial\rho\partial S}}\nabla S -\rho {\frac{\partial E}{\partial S}}\nabla
S
$$
$$
-\Omega ^{\alpha\beta }\nabla\xi _\beta ({\bf v}\nabla )\xi _\alpha
+\nabla\eta ({\bf v}\nabla )S-\nabla S({\bf v}\nabla )\eta =0 \eqno (3.30)
$$
Using (3.28), (3.29) the expression (3.30) reduces to
$$
\partial _0{\bf v}+\nabla {\frac{{\bf v}^2}{2}}+ {\frac{1}{\rho }}%
\nabla(\rho ^2{\frac{\partial E}{\partial\rho }}) -{\bf v}\times
(\nabla\times {\bf v})=0 \eqno (3.31)
$$
In virtue of the identity
$$
{\bf v}\times (\nabla\times {\bf v})\equiv \nabla {\frac{{\bf v}^2}{2}}-(%
{\bf v}\nabla ){\bf v} \eqno (3.32)
$$
the last equation is equivalent to (1.2).

Thus, differentiating equations (3.25), (3.26) and eliminating the variables
$\varphi$, $\bxi$, $\eta$, one obtains the curtailed system (1.1)--(1.3),
whereas the system (1.4), (1.12), (1.13) follows from the system
(3.21)--(3.26) directly (i.e. without differentiating). It means that the
system (1.4), (1.12), (1.13) is an integrated system, whereas the curtailed
system (1.1)--(1.3) is not, although formally they have the same order.

The action of the form (3.20), or close to this form was obtained by some
authors (Seliger and Whithem, 1967; Salmon, 1988), but the quantities $%
g^\alpha $, $\alpha =1,2,3$ are always considered as additional dependent
variables (but not as indefinite functions of $\bxi$ which can be expressed
via initial conditions). The action was not considered as a functional of
fixed indefinite functions $g^\alpha (\bxi)$.

The variable $\eta $ was introduced, for the action be invariant with
respect to the transformations of the whole relabeling group (1.16). To
understand what the $\eta $ means from the mathematical viewpoint, let us
return to the action (3.9), where the internal energy $E$ has the form
(3.10). Adding new variables $j$ by means of designations (2.16), one
obtains instead of (3.14)
$$
{\cal A}_{{\rm E}}[\rho ,{\bf j},\bxi ,p]=\int \{{\frac{{\bf j}^2}{2\rho }}%
-\rho E-p_k[j^k-m\rho _0(\bxi ){\frac{\partial J}{\partial \xi _{0,k}}}]\}dtd%
{\bf x}\eqno (3.33)
$$
where $E$ has the form (3.10).

Variation of (3.33) with respect to $\xi _\alpha$ leads to the equation
$$
\hat{{\cal L}} _\alpha p= \rho {\frac{\partial E(\rho ,S_0(\bxi ))}{\partial
S_0}} {\frac{\partial S_0}{\partial\xi _\alpha}}, \qquad \alpha =1,2,3 \eqno %
(3.34)
$$
where linear operators $\hat{{\cal L}} _\alpha$ are defined by (3.15).
Equations (3.34) are linear non-uniform equations for the variables $p$. A
solution of (3.34) is a sum of the general solution (3.19) of the uniform
equations (3.15) and of a particular solution the non-uniform equations
(3.34). This particular solution depends on the form of the function $S_0$
and cannot be found in a general form. Adding an extraterm $-\eta
j^k\partial _kS$ with $\eta$ satisfying (3.24) to (3.13), a reduction of
non-uniform equations (3.34) to uniform equations (3.15) appears to be
possible. Thus, the extravariable $\eta$ is responsible for the particular
solution of (3.34).

From the viewpoint of the action (3.33) a dependence of the internal energy $%
E$ on the entropy means simply a dependence of $E$ on the labels $\bxi$ via
a function $S(\bxi )$. If such a dependence cannot be expressed through one
function (for instance $E=E[\rho , S_1(\bxi ),S_2(\bxi )]$) the ideal fluid
is described by two entropies $S_1$ and $S_2$ and by two temperatures $%
T_1=\partial E/\partial S_1,\quad T_2=\partial E/\partial S_2$. Such a
situation may appear for a conducting fluid in a strong magnetic field,
where there are two temperatures -- longitudinal and transversal.

Thus five equations (1.4), (1.12), (1.13) with $S$, $\rho $ and ${\bf v}$,
defined respectively by (1.9), (1.10) and (1.11), constitute the fifth order
system for five dependent variables $\xi =\{\xi _{0},\bxi \}$, $\eta$.
Equations (1.1), (1.3),(1.4), (1.12),(1.13) constitute the seventh order
system for seven variables $\rho $, $\bxi $, $\varphi $, $\eta$, $S$.

\medskip

\section{Initial and Boundary Conditions}

Boundary conditions describing vessel walls can be taken into account by
means of a proper choice of the internal energy $E(x,\rho ,S)$ which can
include the energy of the fluid in an external potential $U$.
$$
E=E_{0}(\rho , S)+ U(t,{\bf x}), \eqno (4.1)
$$
\noindent where $U$ is some given external potential. For instance, let the
fluid move inside a volume ${\cal V}$. Then
$$
U({\bf x})=\left\{
\begin{array}{rcc}
0, & \hbox{  inside } & {\cal V} \\
\infty, & \hbox{   outside } & {\cal V} \\
\end{array}
\right.
$$
\noindent Such a choice of the energy $E$ provides that the fluid does not
escape the volume ${\cal V}$.

Let us consider the case, when the fluid flow is considered in the
space-time region $\Omega $ defined by inequalities
$$
\Omega :\qquad t\ge 0,\qquad x^{3}\ge 0 \eqno (4.2)
$$
\noindent The region $\Omega $ has two boundaries: ${\cal I}$ defined by the
relations $t=0$, $x^{3}\ge 0$, and ${\cal B}$ defined by the relations $%
x^{3}=0$, $t\ge 0$. The initial conditions for the system of equations
(1.1)--(1.4) have the form
$$
\rho (0,{\bf x})=\rho _{{\rm in}}({\bf x}),\qquad v^{\alpha }(0,{\bf x})=
v^\alpha _{{\rm in}}({\bf x}),\qquad \alpha =1,2,3 \eqno (4.3)
$$
$$
S(0,{\bf x})=S_{{\rm in}}({\bf x}),\qquad \xi _{\alpha }(0,{\bf x})= \xi
^{\alpha }_{{\rm in}}({\bf x}),\qquad \alpha =1,2,3 \eqno (4.4)
$$
at ${\bf x}\in {\cal I}$ ($t=0$, $x^{3}\ge 0$). Here $\rho _{{\rm in}}$, $%
{\bf v}_{{\rm in}}$, $S_{{\rm in}}$, $\bxi _{{\rm in}}$ are given functions
of argument ${\bf x}$. The boundary conditions on the boundary ${\cal B}$ of
$\Omega $ have the form:
$$
\left.\rho (x)\right|_{ x^{3}=0}=\rho _{{\rm b}}(t,{\bf y}),\qquad \left.
S(x)\right|_{ x^{3}=0}= S_{{\rm b}}(t,{\bf y}),\qquad \{t,{\bf y\}\in {\cal B%
}} \eqno (4.5)
$$
$$
\left. v^{\alpha }(x)\right|_{ x^{3}=0}=v^{\alpha }_{{\rm b}}(t,{\bf y}%
),\qquad \alpha =1,2,3, \qquad\{t,{\bf y\}\in {\cal B}} \eqno (4.6)
$$
$$
\left.\xi _{\alpha }(x)\right|_{ x^{3}=0}=\xi ^{\alpha }_{{\rm b}}(t,{\bf y}%
),\qquad \alpha =1,2,3, \qquad\{t,{\bf y\}\in {\cal B}} \eqno (4.7)
$$
\noindent where
$$
{\bf y\equiv \{}x^{1},x^{2}\} \eqno (4.8)
$$
\noindent Here $\rho _{{\rm b}}$, $S_{{\rm b}}$, ${\bf v}_{{\rm b}}$, $\bxi %
_{{\rm b}}$ are given functions of the argument $\{t,{\bf y}\}$.

Let us show that indefinite functions ${\bf g}$, $S_0$, $\rho _0$ can be
expressed via initial and boundary conditions (4.3)--(4.7). The initial
conditions for the system (3.21)--(3.26) have the form
$$
\xi _\alpha (0,{\bf x})=\xi _{{\rm in}}^\alpha ({\bf x}),\qquad \alpha =1,2,3%
\eqno (4.9)
$$
$$
\rho (0,{\bf x})=\rho _{{\rm in}}({\bf x}),\qquad S(0,{\bf x})=S_0[{\bxi }_{%
{\rm in}}({\bf x})],\eqno (4.10)
$$
$$
\varphi (0,{\bf x})=\varphi _{{\rm in}}({\bf x}),\qquad \eta (0,{\bf x}%
)=\eta _{{\rm in}}({\bf x}),\eqno (4.11)
$$
\noindent (4.9)-(4.11) take place at ${\bf x}\in {\cal I}$. The functions $%
\varphi _{{\rm in}}({\bf x}),\eta _{{\rm in}}({\bf x})$ as well $g^\alpha (%
\bxi )$ are to be determined from the relations
$$
\partial _\alpha \varphi _{{\rm in}}({\bf x})+g^\beta [{\bxi }_{{\rm in}}(%
{\bf x})]\partial _\alpha \xi _{{\rm in}}^\beta ({\bf x})-\eta _{{\rm in}}(%
{\bf x}){\frac{\partial S_0[{\bxi }_{{\rm in}}({\bf x})]}{\partial \xi _{%
{\rm in}}^\beta }}\partial _\alpha \xi _{{\rm in}}^\beta ({\bf x})=
$$
$$
=v_{{\rm in}}^\alpha ({\bf x}),\qquad \alpha =1,2,3;\qquad {\bf x}\in {\cal I%
}.\eqno (4.12)
$$
\noindent It is clear that five functions ${\bf g},\varphi _{{\rm in}},\eta
_{{\rm in}}$ cannot be determined unambiguously from three relations (4.12).

There are at least two different approaches to determination of functions $%
\bxi _{{\rm in}} ({\bf x})$ and ${\bf g}(\bxi )$.

(1) One fixes the functions $\xi ^{\alpha }_{{\rm in}}({\bf x})$ in some
conventional way, sets
$$
\varphi _{{\rm in}}({\bf x})=0,\qquad \eta _{{\rm in}}({\bf x})=0,\qquad
{\bf x}\in {\cal I} \eqno (4.13)
$$

\noindent and determines functions ${\bf g}$ from three relations (4.12).

(2) Functions ${\bf g}$ are fixed in some conventional way, and remaining
functions are determined from relations (4.12)

{\it The first way.} Let the condition (4.9) be given in the form
$$
\xi _{\alpha }(0,{\bf x})=\xi ^{\alpha }_{{\rm in}}({\bf x})=x^{\alpha },
\qquad \alpha =1,2,3,\qquad {\bf x}\in {\cal I}. \eqno (4.14)
$$
\noindent In other words, at $t=0$ the labels $\bxi$ coincide with the
Eulerian coordinates to within a constant factor. The relations (4.12) take
the form
$$
g^{\beta }[{\bxi }_{{\rm in}}({\bf x})]=v^{\beta }_{{\rm in}}({\bf x}),
\qquad \alpha =1,2,3;\qquad {\bf x}\in {\cal I}, \eqno (4.15)
$$
which are resolved in the form
$$
g^{\alpha }(\bxi )=v^{\alpha }_{{\rm in}}(\bxi ),\qquad \alpha =1,2,3,
\qquad\xi _3>0, \eqno (4.16)
$$
\noindent Thus, the functions ${\bf g}$ are expressed through initial
conditions (4.3).

The boundary conditions for the system of equations (3.21)-(3.26) have the
form

$$
\left.\xi _{\alpha }(x)\right|_{ x^{3}=0}=\xi ^{\alpha }_{{\rm b}}(t,{\bf y}%
),\qquad \alpha =1,2,3,\qquad \{t,{\bf y\}\in {\cal B}} \eqno (4.17)
$$
$$
\left. S(x)\right|_{ x^{3}=0}=S_{0}[\bxi _{{\rm b}}(t,{\bf y})]=S_{{\rm b}%
}(t,{\bf y}), \qquad \{t,{\bf y\}\in {\cal B}}, \eqno (4.18)
$$
$$
\left.\rho (x)\right|_{ x^{3}=0}= \left.\rho _{{\rm b}}(x)\right|_{ x^3=0},
\qquad \left. {\bf v}(x)\right|_{ x^{3}=0}= {\bf v}_{{\rm b}}(t,{\bf y}),
\qquad \{t,{\bf y\}\in{\cal B}}, \eqno (4.19)
$$
$$
\left.\varphi (x)\right|_{ x^{3}=0}=\left.\eta (x)\right|_{ x^{3}=0}=0,
\qquad \{t,{\bf y\}\in {\cal B}}, \eqno (4.20)
$$
Let us set
$$
\xi ^{\alpha }_{{\rm b}}(t,{\bf y})=x^{\alpha },\qquad \alpha =1,2;\qquad
\xi ^{3}_{{\rm b}}(t,{\bf y})=-ct,\qquad (t,{\bf y})\in {\cal B}, \eqno %
(4.21)
$$
\noindent where $c$ is a constant.

Writing relations (1.4) and (3.26) for $\xi _{3}<0$ on the boundary ${\cal B}
$ and using (4.20), (4.21), one obtains constraints for the functions ${\bf g%
}(\bxi )$
$$
g^{\beta }[\bxi _{{\rm b}}(t,{\bf y})]\partial _{\alpha }\xi ^{\beta }_{{\rm %
b}}(t,{\bf y})=v^{\alpha }_{{\rm b}}(t,{\bf y}),\qquad \alpha =1,2,\qquad
\{t,{\bf y\}\in {\cal B}} \eqno (4.22)
$$
$$
g^{\beta }[\bxi _{{\rm b}}(t,{\bf y})]\partial _{0}\xi ^{\beta }_{{\rm b}}
(t,{\bf y})=-K_{{\rm b}}(t,{\bf y}),\qquad \{t,{\bf y\}\in {\cal B}}, \eqno %
(4.23)
$$
\noindent where
$$
K_{{\rm b}}(t,{\bf y})\equiv {\frac{{\bf v}^{2}_{{\rm b}}(t,{\bf y})}{2}}+ {%
\frac{\partial \{\rho _{{\rm b}}(t,{\bf y})E[\rho _{{\rm b}}(t,{\bf y}), S_{%
{\rm b}}(t,{\bf y})]\}}{\partial \rho _{{\rm b}}(t,{\bf y})}}, \qquad\{t,%
{\bf y\}\in {\cal B}}, \eqno (4.24)
$$

Substituting relations (4.21) into (4.22), (4.23), one obtains three
equations for determination of functions ${\bf g}(\bxi )$. Resolving this
system of equations with respect to ${\bf g}$, one obtains
$$
g^\alpha (\bxi )=v_{{\rm b}}^\alpha (-\xi _3/c,\xi _1,\xi _2),\qquad \alpha
=1,2;\qquad \xi _3<0
$$
$$
g^3(\bxi )=c^{-1}K_{{\rm b}}(-\xi _3/c,\xi _1,\xi _2),\qquad \xi _3<0\eqno %
(4.25)
$$
Thus, ${\bf g}(\bxi )$ is determined by (4.16) for $\xi _3>0$ and by (4.25)
for $\xi _3<0$. In other words, the boundary conditions and the initial
conditions determine the vector field ${\bf g}(\bxi )$ in different regions
of the argument $\bxi$. All information about the velocities at the initial
moment and on the boundary has been transferred into dynamic equations. The
field ${\bf g}(\bxi )$ can describe both initial and boundary conditions.

{\it The second way}. Let us choose the functions ${\bf g}$ in a simple
form. Let for instance,
$$
g^{1}(\bxi )=\xi _{2},\qquad g^{2}(\bxi )=0,\qquad g^{3}(\bxi )=0 \eqno %
(4.26)
$$
\noindent Let us set
$$
\chi =\varphi ,\qquad \lambda =\xi _2,\qquad \mu =\xi _1\eqno (4.27)
$$
\noindent Then the expression (1.11) takes the form
$$
{\bf u}(\chi ,\lambda ,\mu ,\eta ,S)\equiv \nabla \chi +\lambda \nabla \mu
-\eta \nabla S={\bf v}\eqno (4.28)
$$
\noindent where $\chi ,\lambda ,\mu $, are Clebsch potentials (Clebsch,
1857; 1859). Now six equations (1.1), (1.3), (3.22)-(3.26), (3.27) [(3.22)
for $\alpha =3$ is of no importance] for six dependent variables $\rho ,\chi
,\lambda ,\mu ,\eta ,S$ do not contain indefinite functions and have an
unambiguous form.
$$
\partial _0\rho +\nabla (\rho {\bf u})=0,\qquad \partial _0\lambda +({\bf u}%
\nabla )\lambda =0
$$
$$
\partial _0\mu +({\bf u}\nabla )\mu =0,\qquad \partial _0S+({\bf u}\nabla
)S=0\eqno (4.29)
$$
$$
\partial _0\eta +({\bf u}\nabla )\eta =-{\frac{\partial E}{\partial S}}%
,\qquad \partial _0\chi +\lambda \partial _0\mu -\eta \partial _0S+{\frac 12}%
{\bf u}^2+{\frac{\partial (\rho E)}{\partial \rho }}=0
$$
where ${\bf u}$ is defined by (4.28).

The initial conditions for variables $\rho ,\chi , \lambda , \mu , \eta , S$
are determined by relations
$$
\rho (0,{\bf x})=\rho _{{\rm in}}(0,{\bf x}),\qquad S(0,{\bf x})=S_{{\rm in}%
}(0,{\bf x}),\eqno (4.30)
$$
$$
\nabla \chi _{{\rm in}}+\lambda _{{\rm in}}\nabla \mu _{{\rm in}}-\eta _{%
{\rm in}}\nabla S_{{\rm in}}={\bf v}_{{\rm in}}\eqno (4.31)
$$
\noindent Three equations (4.30), (4.31) do not determine the initial
conditions
$$
\chi (0,{\bf x})=\chi _{{\rm in}}({\bf x}),\qquad \lambda (0,{\bf x}%
)=\lambda _{{\rm in}}({\bf x}),\eqno (4.32)
$$
$$
\mu (0,{\bf x})=\mu _{{\rm in}}({\bf x}),\qquad \eta (0,{\bf x})=\eta _{{\rm %
in}}({\bf x}),\eqno (4.33)
$$
\noindent unambiguously.

If the fluid is described in terms of Clebsch potentials, the dynamic
equations contain neither arbitrary functions, nor information about the
initial conditions. It should be interpreted in the sense that the
description (4.28)-(4.29) in terms of the Clebsch potentials is a result of
a change of variables in dynamic equations (1.1)-(1.3), whereas the
description (3.21)-(3.26) is a result of integration of the dynamic
equations (1.1)-(1.4). In other words, the description (4.28)-(4.29) in
terms of Clebsch potentials relates to the description (3.21)-(3.26) in the
same way, as a particular solution of a system of differential equations
relates to a general solution of the same system.

Let us note that there are many other ways for determination of indefinite
functions ${\bf g}(\bxi )$. For instance, for slightly rotational flows the
functions ${\bf g}(\bxi )$ may be chosen as small corrections to the basic
irrotational flow described by the potential $\varphi$.

\section{Description in Lagrangian coordinates}

To show that the system (1.4), (1.9)-(1.13) is indeed the integrated system
of hydrodynamic equations, let us rewrite it in Lagrangian coordinates, when
five variables ${\bf x}=\{x^1,x^2,x^3\},\varphi ,\eta $ are considered as
functions of four independent variables $t,\bxi =\{\xi _1,\xi _2,\xi _3,\}$.
It is necessary to introduce designations
$$
Q\equiv {\frac{\partial (x^1,x^2,x^3)}{\partial (\xi _1,\xi _2,\xi _3)}}%
\equiv {\frac{\partial ({\bf x})}{\partial (\bxi )}}\equiv \det \parallel
x^{\alpha ,\beta }\parallel ,\qquad x^{\alpha ,\beta }\equiv {\frac{\partial
x^\alpha }{\partial \xi _\beta }},\qquad \alpha ,\beta =1,2,3\eqno (5.1)
$$
$$
X_{\alpha ,\beta }\equiv {\frac{\partial Q}{\partial x^{\alpha ,\beta }}}%
,\qquad \alpha ,\beta =1,2,3\eqno (5.2)
$$
\noindent It follows from (2.3), (3.4) that
$$
Q=({\frac{\partial J}{\partial \xi _{0,0}}})^{-1}={\frac{m\rho _0(\bxi )}%
\rho }.\eqno (5.3)
$$
\noindent Identities (2.7), (2.8) take the form
$$
x^{\alpha ,\beta }X_{\gamma ,\beta }\equiv \delta _\gamma ^\alpha Q,\qquad {%
\frac \partial {\partial \xi _\gamma }}X_{\alpha ,\gamma }\equiv 0,\qquad
\alpha ,\beta =1,2,3.\eqno (5.4)
$$
\noindent Derivative with respect $x^1 $ can be recalculated into derivative
with respect to $\xi _\alpha $ as follows
$$
{\frac{\partial \varphi }{\partial x^1}}\equiv {\frac{\partial (\varphi
,x^2,x^3)}{\partial (x^1,x^2,x^3)}}\equiv {\frac{\partial (\varphi ,x^2,x^3)%
}{\partial (\xi _1,\xi _2,\xi _3)}}{\frac{\partial (\xi _1,\xi _2,\xi _3)}{%
\partial (x^1,x^2,x^3)}}=Q^{-1}X_{1,\beta }{\frac{\partial \varphi }{%
\partial \xi _\beta }}
$$
\noindent or for a derivative with respect to $x^\alpha $
$$
{\frac{\partial \varphi }{\partial x^\alpha }}\equiv Q^{-1}X_{\alpha ,\beta }%
{\frac{\partial \varphi }{\partial \xi _\beta }}\qquad \alpha =1,2,3\eqno %
(5.6)
$$
\noindent Applying the rule (5.6) to (1.11), one obtains
$$
v^\alpha =v^\alpha (t,\bxi )=Q^{-1}X_{\alpha ,\beta }[{\frac{\partial
\varphi }{\partial \xi _\beta }}+g^\beta (\bxi )-\eta {\frac{\partial S_0(%
\bxi )}{\partial \xi _\beta }}],\qquad \alpha =1,2,3\eqno (5.7)
$$
\noindent Let $D/Dt$ means derivative with respect to $t$ at constant $\bxi $
$$
{\frac{D\varphi }{Dt}}\equiv \partial _0\varphi +{\frac{Dx^\alpha }{Dt}}%
\partial _\alpha \varphi \eqno (5.8)
$$
\noindent In particular
$$
{\frac{D\xi _\alpha }{Dt}}=\partial _0\xi _\alpha +{\frac{Dx^\beta }{Dt}}%
\partial _\beta \xi _\alpha =0,\eqno (5.9)
$$
\noindent because $D\xi _\alpha /Dt=0$ by definition. Comparing (5.9) with
(1.4) and using (5.7), one concludes that
$$
v^\alpha ={\frac{Dx^\alpha }{Dt}}=Q^{-1}X_{\alpha ,\beta }[{\frac{\partial
\varphi }{\partial \xi _\beta }}+g^\beta (\bxi )-\eta {\frac{\partial S_0(%
\bxi )}{\partial \xi _\beta }}],\qquad \alpha =1,2,3\eqno (5.10)
$$
\noindent Now in virtue of (5.6), (5.7), (5.10) the equations (1.12), (1.13)
can be rewritten in the form
$$
{\frac{D\varphi }{Dt}}-{\frac 12}\sum_{\alpha =1}^{\alpha =3}\left\{
Q^{-1}X_{\alpha ,\beta }\left[ {\frac{\partial \varphi } {\partial \xi
_\beta }}+g^\beta (\bxi )-\eta {\frac{\partial S_0(\bxi )} {\partial \xi
_\beta }}\right]\right\}^2+{\frac{\partial [\rho E(\rho ,S)]}{\partial \rho }%
}=0\eqno (5.11)
$$
$$
{\frac{D\eta }{Dt}}=-{\frac{\partial E(\rho ,S)}{\partial S}}=0\eqno (5.12)
$$
\noindent where
$$
S=S_0(\bxi ),\qquad \rho ={\frac{m\rho _0(\bxi )}Q},\qquad Q={\frac{\partial
({\bf x})}{\partial (\bxi )}}.\eqno (5.13)
$$
\noindent The system of five equations (5.10)-(5.12) is a system for five
dependent variables ${\bf x},\varphi ,\eta $, considered as functions of
four independent variables $t,\bxi $. Five indefinite functions ${\bf g}%
,\rho _0,S_0$ of $\bxi $ are determined from initial and boundary
conditions. In particular, if there is no inflow of the fluid and the
initial values for ${\bf x},\varphi ,\eta $ are given in the form
$$
{\bf x}(0,\bxi )=\bxi ,\eqno (5.14)
$$
$$
\varphi (0,\bxi )=0,\qquad \eta (0,\bxi )=0,\eqno (5.15)
$$
\noindent it follows from (5.10), (5.13), that
$$
S(0,\bxi )=S_0(\bxi ),\qquad \rho (0,\bxi )=\rho _0(\bxi ),\qquad {\frac{D%
{\bf x}}{Dt}}(0,\bxi )={\bf v}_{{\rm in}}(\bxi )={\bf g}(\bxi )\eqno (5.16)
$$
In the Lagrangian description the initial conditions (5.14) for the position
of a particle labeled by $\bxi $ look quite reasonable. From physical point
of view a necessity of initial conditions for the particle position does not
raise doubts. The system of hydrodynamic equations (5.10)-(5.12) in the
Lagrangian form looks as partly integrated system. Indeed, the system
(1.1)-(1.4) written in the independent Lagrangian coordinates $\bxi $ for
six dependent variables ${\bf x}={\bf x}(t,\bxi ),{\bf v}={\bf v}(t,\bxi )$
has the form
$$
{\frac{Dx^\alpha }{Dt}}=v^\alpha ,\qquad m{\frac{Dv^\alpha }{Dt}}=-\rho _0(%
\bxi )^{-1}X_{\alpha ,\beta }{\frac \partial {\partial \xi _\beta }}{\frac
\partial {\partial \rho }}[\rho ^2E(\rho ,S_0(\bxi ))],\qquad \alpha =1,2,3%
\eqno (5.17)
$$
\noindent where $X_{\alpha ,\beta }$ and $\rho $ are functions of $\bxi$, $%
x^\alpha $ and $x^{\alpha ,\beta }\equiv \partial x^\alpha /\partial \xi
^\beta$ which are defined by relations (5.1), (5.2), (5.13). The six order
system (5.17) contains only two indefinite functions $\rho _0(\bxi )$ and $%
S_0(\bxi )$ describing initial values of density and entropy. Initial values
of velocity ${\bf v}(0,\bxi )$ and position ${\bf x}(0,\bxi )$ are given by
initial conditions
$$
{\bf x}(0,\bxi )=\bxi ,\qquad {\bf v}(0,\bxi )={\bf v}_{{\rm in}}(\bxi )%
\eqno (5.18)
$$
The relation (5.7) is an integral of (5.17). It satisfies the equations
(5.17) for any functions ${\bf g}$ in virtue of equations (5.10)-(5.12),
although this circumstance is not evident directly.

Note that the curtailed system (1.1)-(1.3) cannot be written in the
Lagrangian form directly, because it does not contain any reference to
Lagrangian coordinates $\bxi $. To introduce $\bxi $, it is necessary to
append equations (1.4). Then fifth order system (1.1)-(1.3) turns to the
complete eight order system (1.1)-(1.4). Equations (1.1) and (1.3) can be
integrated on the basis of (1.4) in the form (1.9), (1.10). The remaining
equations (1.2), (1.4) constitute the sixth order system which can be
written in the Lagrangian form (5.17).

\section{Incompressible fluid}

In the special case of the incompressible fluid, it should set $\rho =\rho
_0=$const in the action (3.20) and introduce new variable
$$
{\bf v}={\bf j}/\rho _0, \qquad \rho _0={\rm const} \eqno (6.1)
$$
It is easy to verify that $\eta =\eta (\bxi )$, $S=S_0(\bxi )$, and the last
term of (3.20) can be incorporated in the term $j^kg^\alpha (\bxi )\partial
_k\bxi _\alpha$. Thus, the action for the incompressible fluid looks as
follows
$$
{\cal A}_{{\rm E}}[{\bf v},\bxi ,\varphi ]=\rho _0\int \{ {\frac{{\bf v}^{2}
}{2}}-{\bf v}\nabla\varphi-g^\alpha (\bxi )\partial _0\xi _\alpha -g^\alpha (%
\bxi ){\bf v}\nabla\xi _\alpha \}dtd{\bf x}, \eqno (6.2)
$$
where $g^\alpha (\bxi )$ are arbitrary fixed functions of $\bxi$.

Variation with respect to ${\bf v}$, $\bxi$, $\varphi$ gives
$$
\delta {\bf v}:\qquad {\bf v}=\nabla\varphi +g^\alpha (\bxi ) \nabla\xi
_\alpha \eqno (6.3)
$$
$$
\rho _0^{-1}{\frac{\delta {\cal A}_{{\rm E}}}{\delta\xi _\alpha }}=
(g^{\alpha ,\beta }-g^{\beta ,\alpha })(\partial _0\xi _\alpha + {\bf v}%
\nabla\xi _\alpha)=0, \qquad \alpha =1,2,3 \eqno (6.4)
$$
$$
\rho _0^{-1}{\frac{\delta {\cal A}_{{\rm E}}}{\delta\varphi }}= \nabla {\bf v%
}=0 \eqno (6.5)
$$

In the general case the condition (3.27) is satisfied. Substituting (6.3)
into (6.4) and (6.5), one obtains
$$
\partial _{0}\xi _{\alpha }+[\nabla \varphi +g^{\beta }(\bxi )\nabla \xi
_{\beta }]\nabla \xi _{\alpha }=0,\qquad\alpha =1,2,3 \eqno (6.6)
$$
$$
\nabla ^{2}\varphi +g^{\alpha ,\beta }(\bxi )\nabla \xi _{\beta }\nabla \xi
_{\alpha }+g^\alpha (\bxi )\nabla ^2\xi _\alpha =0 \eqno (6.7)
$$
\noindent The dynamic equation for $\varphi $ does not contain temporal
derivative.

\noindent Conventional hydrodynamic equations for the incompressible fluid
$$
\nabla {\bf v}=0,\qquad \partial _0{\bf v}+ ({\bf v}\nabla ){\bf v}=-{\frac{%
\nabla p}{\rho _{0}}} \eqno (6.8)
$$
\noindent are obtained from relations (6.3)-(6.5). Differentiating (6.3)
with respect to $t$, one obtains
$$
\partial _0{\bf v}=\nabla [\partial _{0}\varphi +g^{\alpha }(\bxi )\partial
_0\xi _{\alpha }] -\Omega ^{\alpha\beta }\partial _0\xi _\beta\nabla\xi
_\alpha \eqno (6.9)
$$
\noindent where $\Omega ^{\alpha\beta }$ is defined by (1.21). It follows
from (6.3), (3.28),(3.29)
$$
{\bf v\times }(\nabla \times {\bf v})=\Omega ^{\alpha \beta }(\bxi ) \nabla
\xi _{\beta }({\bf v\nabla })\xi _{\alpha }. \eqno (6.10)
$$
\noindent In virtue of (6.4) the last term in rhs of (6.9) coincides with
rhs of (6.10). Then using the identity (3.32), one obtains
$$
\partial _0{\bf v}+({\bf v\nabla }){\bf v}= \nabla [\partial _{0}\varphi +g
^\alpha (\bxi)\partial _0\xi _\alpha +{\frac{1}{2}}{\bf v}^{2}] \eqno (6.11)
$$
The equation (6.11) coincides with the second equation (6.8), provided one
uses designation
$$
{\frac{p}{\rho _{0}}}=-{\frac{1}{2}}{\bf v}^{2}-\partial _{0}\varphi
-g^\alpha (\bxi )\partial _0\xi _\alpha \eqno (6.12)
$$
\noindent Here the pressure $p$ is determined after solution of the system
of hydrodynamic equations (6.3)-(6.5), or (6.8).

If the condition (3.27) is fulfilled, the equation (6.4) may be written in
the form (6.6). Then eliminating $\partial _0\xi _\alpha $ by means of (6.6)
and using the corollary of (6.3)
$$
g^\alpha (\bxi )({\bf v\nabla })\xi _\alpha ={\bf v}^2-({\bf v\nabla }%
)\varphi ,\eqno (6.13)
$$
the relation (6.12) for the pressure can be written in the form
$$
{\frac p{\rho _0}}={\frac 12}{\bf v}^2-[\partial _0\varphi +({\bf v\nabla }%
)\varphi ]\eqno (6.14)
$$

In the case of a irrotational flow, when ${\bf v}=\nabla \varphi $, the
inequality (3.27) turns to equality and formally the derivation of the
expression (6.13) is not founded. Nevertheless the relation (6.14) remains
valid in this case, because it turns to the integral
$$
{\frac p{\rho _0}}+{\frac 12}{\bf v}^2+\partial _0\varphi =0\eqno (6.15)
$$
The relation (6.14) is valid for any flow of incompressible fluid, but it is
not an integral. It is a definition. From point of view of the description
in terms of hydrodynamic potentials $\varphi $, $\bxi$, the relation (6.14)
is a definition of the pressure $p$ in terms of dependent variables $\varphi
$, $\bxi$ and the relation (6.3). From the viewpoint of the curtailed system
(6.8) the relation (6.14) is a definition of the function $\varphi $ in
terms of variables ${\bf v}$, $p$. Nevertheless, this definition is useful
for a description of a slightly rotational flow, when the velocity ${\bf v}$
can be represented in the form
$$
{\bf v}=\nabla \varphi +\delta {\bf v},\qquad \mid \delta {\bf v}\mid \ll
\mid \nabla \varphi \mid ,\qquad \delta {\bf v}=g^\alpha (\bxi )\nabla \xi
_\alpha ,\eqno (6.16)
$$
where $\delta {\bf v}$ describes a small rotational component of the
velocity.

Substituting (6.12) into (6.14), one obtains
$$
{\frac{p}{\rho _{0}}}=-\partial _{0}\varphi -{\frac{1}{2}}(\nabla\varphi )^2
+{\frac{1}{2}} (\delta {\bf v})^{2} \eqno (6.17)
$$
This relation shows that a contribution of the small rotational component $%
\delta {\bf v}$ into the pressure is of the second order $(\delta {\bf v}%
)^{2}$ and this contribution always increases the pressure.

\section{Slightly Rotational Flow of Incompressible Fluid}

It seems to be reasonable to consider a slightly rotational flow as a small
correction to an eatablished irrotational flow which can be effectively
calculated for a flow around different bodies. Let $u$ be a set of dependent
variables, and dependent variables $u_0$ describe an irrotational flow. Let
us represent a slightly rotational flow in the form
$$
u=u_0+\varepsilon u_1+O(\varepsilon ^2),\qquad \varepsilon \ll 1 \eqno (7.1)
$$
where $\varepsilon $ is a small formal parameter which is set to be equal to
1 after calculation. It means that the quantities $u_1$ are considered as
small with respect to $u_0$, $u_1/u_0=O(\varepsilon )\ll 1$. Substituting
(7.1) into dynamic equations and neglecting higher order terms, one obtains
linear equations for $u_1$. Use of dependent variables ${\bf v}$, $p$,
satisfying equations (6.8), leads to the following first order approximation
equations.
$$
\nabla {\bf v}_1=0,\qquad \partial _0{\bf v}_1+(\nabla \varphi _0\nabla )
{\bf v}_1+ ({\bf v}_1\nabla )\nabla \varphi _0=-\nabla p_1/\rho _0, \qquad
{\bf v}_0=\nabla\varphi _0 \eqno (7.2)
$$
Dynamic equation for the rotational component $\bomega _1=\nabla \times {\bf %
v}_1$ of the velocity follows from the second equation (7.2)
$$
\partial _0\bomega _1+(\nabla \varphi _0\nabla )\bomega _1- (\bomega %
_1\nabla )\nabla\varphi _0=0 \eqno (7.3)
$$
Both equations (7.2) and (7.3) are difficult for a solution. Use of dynamic
equations (6.3), (6.6) and (6.7) for hydrodynamic potentials appears to be
more effective, because for a fixed velocity ${\bf v}=\nabla\varphi _0$
equations (1.4) for $\bxi$ are equivalent to the system of ordinary
equations (1.6) which can be solved simply enough. Indeed, let the velocity $%
{\bf v}=\nabla\varphi_0$ of the basic established irrotational flow be
known, and the orthogonal coordinate system $\xi, \eta, \zeta$ coupled with
the flow be determined by the relations
$$
\xi =u_\infty ^{-1}\varphi _0({\bf x}),\qquad \nabla \xi \cdot \nabla \eta
=0, \qquad \nabla \eta \cdot \nabla \zeta =0,\qquad \nabla \xi \cdot \nabla
\zeta =0\eqno (7.4)
$$
where $u_\infty$ is a constant velocity of the flow at infinity. Then the
general solution of the equations (1.4) is written in the form
$$
\xi _\alpha =f_\alpha (X-u_\infty t,\eta ,\zeta ),\qquad \alpha =1,2,3 \eqno %
(7.5)
$$
$$
X=X(\xi ,\eta ,\zeta )=\int\limits_0^\xi V_0^{-2}(\xi ^{\prime },\eta ,\zeta
)d\xi ^{\prime },\qquad V_0^2=(\nabla \xi )^2={(\nabla \varphi _0/u_\infty )}%
^2 \eqno (7.6)
$$
where $f_\alpha $, $\alpha =1,2,3$ are arbitrary functions of three
arguments $X-u_\infty t$, $\eta $, $\zeta $. $V_0=V_0(\xi ,\eta ,\zeta )$ is
a dimensionless velocity of the basic flow considered as a function of
coordinates $(\xi ,\eta ,\zeta )$. This fact can be tested by means of a
direct substitution of (7.6) into equation (1.4) with ${\bf v}=\nabla\varphi
_0$. $X$ can be interpreted as a ''distorted'' Cartesian coordinate in the
direction along the basic flow at infinity. The distortion of the coordinate
$X$ is chosen in such a way that the flow were uniform.

Let us use the expansion (7.1) for dependent variables $u=\{\varphi ,\bxi \}$
$$
{\bf v=v}_0+\varepsilon {\bf v}_1+O(\varepsilon ^2)=\nabla \left( \varphi
_0+\varepsilon \varphi _1\right) +\varepsilon g^\alpha (\bxi )\nabla \xi
_\alpha +O(\varepsilon ^2)\eqno (7.7)
$$
$$
{\bf v}_0=\nabla \varphi _0,\qquad {\bf v}_1=\nabla\varphi _1+g^\alpha (\bxi %
)\nabla \xi _\alpha \eqno (7.8)
$$

Substituting (7.7) into (6.7), (6.6) and equating coefficients before equal
powers of $\varepsilon $ to zero, one obtains
$$
\varepsilon ^0:\qquad \nabla ^2\varphi _0=0\eqno (7.9)
$$
$$
\varepsilon :\qquad\partial _0\xi _\alpha +\nabla \varphi _0\nabla \xi
_\alpha =0,\qquad \alpha =1,2,3 \eqno (7.10)
$$
$$
\varepsilon :\qquad\nabla ^2\xi _\alpha +\nabla [g^\alpha (\bxi ) \nabla \xi
_\alpha ]=0, \eqno (7.11)
$$
Let us note that the labels $\bxi $ are calculated only in zeroth
approximation. Instead of the expansion of $\bxi $ one uses the supposition
that the rotational components $g^\alpha (\bxi )$ of the velocity are small
as compared with the irrotational component $\nabla \varphi _0$. General
solution of equations (7.10) have the form (7.5). Then (7.11) takes the form
$$
\nabla ^2\varphi _1+\nabla \left( A_\alpha \nabla B_\alpha \right) =0,\qquad
g^\alpha (\bxi )\nabla \xi _\alpha =A_\alpha \nabla B_\alpha \eqno (7.12)
$$
where $A_\alpha $, $B_\alpha $, $\alpha =1,2,3$ are arbitrary functions of
arguments $X-u_\infty t,$ $\eta $, $\zeta $. Let $\Sigma $ be the solid
boundary of the volume with the fluid. Then equation (7.12) should be solved
with the boundary condition
$$
\left. {\bf n}\nabla \varphi _1\right| _\Sigma =-\left. A_\alpha {\bf n}%
\nabla B_\alpha \right| _\Sigma \eqno (7.13)
$$

Note that $\varphi _1$, $\bxi $ are hydrodynamic potentials which are not
determined inambiguously by the fluid flow. But any set of hydrodynamic
potentials determines inambiguously a fluid flow. It is possible one to
express the functions ${\bf A}=\{A{\bf _\alpha \}}$, ${\bf B}=\{B_\alpha \}$%
, $\alpha =1,2,3$ via functions ${\bf g}(\bxi
)=\{g^\alpha (\bxi )\}$, $\alpha =1,2,3$, choosing at $t=0$ $\bxi ={\bf x}$.
Then the relations

$$
{\bf B}(X,\eta ,\zeta )={\bf x},\qquad {\bf A}(X,\eta ,\zeta )={\bf g}({\bf x%
})\eqno (7.14)
$$
determine the form of functions ${\bf A}$, ${\bf B}$ and the fact that ${\bf %
A}$ is a function of ${\bf B}$:\quad ${\bf A}$=${\bf g}({\bf B})$. On the
other hand the functions ${\bf g}(\bxi )$ are indefinite functions whose
form is determined by the initial conditions and by a choice of the labeling
$\bxi $. The relabeling transformation (1.16) changes the form of the
functions ${\bf g}(\bxi )$.

Besides only the rotational component of the vector $A_\alpha \nabla
B_\alpha $ is essential, because its irrotational component is compensated
by the contribution of $\varphi _1$ into ${\bf v}$. Indeed, Let $A_\alpha
\nabla B_\alpha $can be represented in the form

$$
A_\alpha \nabla B_\alpha =\nabla \Phi +\nabla \times {\bf C,\qquad }\nabla
{\bf C}=0\eqno (7.15)
$$
where ${\bf C}$ is some solenoidal vector describing the rotational
component of $A_\alpha\nabla B_\alpha$. Substitution of (7.15) into (7.12),
(7.13) leads to the relations
$$
\nabla ^2\widetilde{\varphi }_1=0,\qquad \widetilde{\varphi }_1=\varphi
_1+\Phi ,\quad \left. {\bf n}\nabla \widetilde{\varphi }_1\right| _\Sigma
=-\left. {\bf n}\nabla \times {\bf C}\right| _\Sigma \eqno (7.16)
$$

In other words, the potential $\varphi _1$ compensates the irrotational
component of $A_\alpha\nabla B_\alpha$. As a result the total potential $\\%
tilde{varphi _1}$ satisfies the same equation (7.9) as $\varphi _0$ does.
The boundary condition for $\tilde{\varphi }_1$ appears to be qenerated by
the rotational component of ${\bf v}_1$
$$
{\bf v}=\nabla (\varphi _0 +\varepsilon\tilde{\varphi }_1) +\nabla\times
{\bf C} +O(\varepsilon ^2), \qquad \nabla ^2(\varphi _0 +\varepsilon\tilde{%
\varphi }_1) =0 \eqno (7.17)
$$
$$
\left. {\bf n}\nabla (\varphi _0 +\varepsilon\tilde{\varphi }_1)\right| _
\Sigma =-\left. \varepsilon {\bf n}\times {\bf C}\right|_\Sigma \eqno (7.18)
$$
Solution of the equation (7.12) with the boundary condition (7.13) is
written in the form:
$$
\varphi _1({\bf x})=\frac 1{4\pi }\int\limits_VG({\bf x,x}^{\prime })\nabla
^{\prime }[A_\alpha ({\bf x}^{\prime })\nabla ^{\prime }B_\alpha ({\bf x}%
^{\prime })]d{\bf x}^{\prime }-\frac 1{4\pi }\oint G({\bf x,x}^{\prime
})A_\alpha ({\bf x}^{\prime })\nabla ^{\prime }B_\alpha ({\bf x}^{\prime })d%
{\bf S}^{\prime }\eqno (7.19)
$$
where $\nabla ^{\prime }$ means the gradient with respect to coordinates $%
{\bf x}^{\prime }$, and $G({\bf x,x}^{\prime })$ is the Green function
satisfying the following conditions
$$
\nabla ^{\prime 2}G({\bf x,x}^{\prime })=-4\pi \delta ({\bf x-x}^{\prime
}),\qquad \left. \frac{\partial G({\bf x,x}^{\prime })}{\partial n^{\prime }}%
\right| _{{\bf x^\prime }\in\Sigma }=0\eqno (7.20)
$$
where $\partial /\partial n$ means component of the gradient in the
direction of the normal, and functions $A_\alpha$, $B_\alpha$ depend only on
arguments $X-u_\infty t$, $\eta$, $\zeta$ which are known functions of ${\bf %
x}$. The surface integral is taken over the surface surrounding the volume $V
$ with the fluid. This surface includes the surface $\Sigma $ of the body
and the infinite surface around $V$. The surface integral in (7.19) contains
functions $A_\alpha ({\bf x}^{\prime })$, $B_\alpha ({\bf x}^{\prime })$
determined in the whole volume $V$. It permits to transform the surface
integral to the volume one
$$
\frac 1{4\pi }\oint G({\bf x,x}^{\prime })A_\alpha ({\bf x}^{\prime })\nabla
^{\prime }B_\alpha ({\bf x}^{\prime })d{\bf S}^{\prime }=\frac 1{4\pi
}\int\limits_V\nabla ^{\prime }[G({\bf x,x}^{\prime })A_\alpha ({\bf x}^
{\prime })\nabla ^{\prime }B_\alpha ({\bf x}^{\prime })]d{\bf x}^{\prime } %
\eqno (7.21)
$$
Then the relation (7.19) transforms to
$$
\varphi _1({\bf x})=-\frac 1{4\pi }\int\limits_V\nabla ^{\prime }G({\bf x,x}%
^{\prime })\cdot A_\alpha ({\bf x}^{\prime })\nabla ^{\prime }B_\alpha ({\bf %
x}^{\prime })d{\bf x}^{\prime } \eqno (7.22)
$$

Thus, the slightly rotational flow is described by the relations
$$
{\bf v}=\nabla \varphi _0+\varepsilon \left( \nabla \varphi _1+A_\alpha
\nabla B_\alpha \right) +O\left( \varepsilon ^2\right) \eqno (7.23)
$$
$$
\frac p{\rho _0}=-\partial _0\left( \varphi _0+\varepsilon \varphi _1\right)
-\frac 12(\nabla \varphi _0)^2-\varepsilon \nabla \varphi _0\nabla \varphi
_1+O\left( \varepsilon ^2\right) \eqno (7.24)
$$
where $A_\alpha $, $B_\alpha $, $\alpha =1,2,3$ are arbitrary functions of
arguments $X-u_\infty t$, $\eta $, $\zeta $ and $\varphi _1$ is described by
the relation (7.22). Expressions for $p$ is a result of application of
(6.17) to the case of the slightly rotational flow.

Relations (7.23), (7.24) satisfy the equations (7.2), (7.3), but a direct
derivation of (7.23), (7.24) from (7.2), (7.3) seems to be rather difficult
because of terms $(\bomega _1\nabla )\nabla \varphi _0$, $( {\bf v}_1\nabla
) \nabla \varphi _0$. Such terms connect three similar equations of the type
(1.4) into united system of three equations and prevent to use equivalence
of one equation of the type (1.4) and the system (1.6) of ordinary equations.

The quantities $A_\alpha (\bxi )$, $B_\alpha (\bxi )$ and other functions of
only $\bxi $, for instance $\Omega ^{\alpha \beta }(\bxi )$, defined by
(1.21), depend on only arguments $X-u_\infty t$, $\eta $, $\zeta $ and can
be regarded as quantities frozen in the basic flow, because they depend on $%
t $ and $X$ only through $X-u_\infty t$. Gradients of frozen quantities, for
instance $A_\alpha \nabla B_\alpha $, are not frozen in the basic flow, in
general, because they depend on $t$ and $X$ not only via argument $%
X-u_\infty t$. They depend on $X$ also through arguments $\nabla \xi \cdot
\partial X/\partial \xi =\nabla \xi /\left| \nabla \xi \right| ^2$, $\nabla
\eta $, $\nabla \zeta $ which depend, in general on $X$, but do not depend
on $t$. It is a reason why a description in terms of ''frozen quantities''
appears to be more effective, than that in terms of the ''frozen quantities
gradients''. Appearance of extraterms in (7.2), (7.3) as compared with (1.4)
is connected with the use of the gradients of frozen quantities.

\section{Wave Function and Kinematic Spin of a Flow}

There exists a special complex form of hydrodynamic potentials. In this form
the dynamic equation for the irrotational component of the flow is very
close to a linear equation.

Idea of the transformation of hydrodynamic potentials is very simple. Let
there be an irrotational flow ${\bf v}=b\nabla \phi $ of a compressible
fluid whose internal energy depends only on $\rho $. Here $b$ is a constant
of dimensionality $[b]=[L^2T^{-1}]$ which is introduced to make the
potential $\phi $ to be dimensionless. Then dynamic equations (1.1), (1.2)
can be integrated and written in the form
$$
\partial _0\rho +b\nabla (\rho\nabla \phi )=0, \qquad b\partial _0\phi +{%
\frac{b^{2}}{2}}(\nabla \phi )^{2} =-{\frac{\partial }{\partial \rho }}[\rho
E(\rho )] \eqno (8.1)
$$
\noindent Equation (1.2) is obtained as a gradient of the second equation
(8.1). The term $b^{2}(\nabla \phi )^{2}/2$ is the principal nonlinear term
known as convective nonlinearity term $({\bf v\nabla })${\bf v}. Introducing
the complex variable $\Psi $, this term can be removed. Let
$$
\Psi =\sqrt{\rho } e^{i\phi },\qquad \rho =\bar{\Psi }\Psi , \qquad {\bf v}=-%
{\frac{ib}{2\bar{\Psi }\Psi }}(\bar{\Psi }\nabla \Psi -\nabla \bar{\Psi }%
\Psi ) \eqno (8.2)
$$

\noindent The bar over symbol means the complex conjugation. Then equations
(8.1) are described in the form of one complex equation for the complex
dependent variable $\Psi $.
$$
ib\partial _0\Psi +{\frac{b^{2}}{2}}\nabla ^2\Psi = \{{\frac{\partial }{%
\partial \rho }}[\rho E(\rho )]+ {\frac{b^{2}}{2\sqrt{\rho }}}\nabla ^{2}(%
\sqrt{\rho })\}\Psi , \qquad \rho \equiv \bar{\Psi }\Psi \eqno (8.3)
$$
\noindent Equations (8.1) are imaginary and real components of (8.3). The
second term in lhs of (8.3) is linear. It corresponds to the convective term
$({\bf v\nabla })${\bf v}. Now this term contains the second order spatial
derivative. It is a price which is paid for the linearity. As a result of
the transformation (8.2) the nonlinearity is transmitted from the kinematic
term to the dynamic one. The kinematic nonlinearity is stronger, than the
dynamic nonlinearity, because the kinematic non-linearity is connected
directly with velocities, whereas the dynamic nonlinearity is coupled
directly only with forces and with accelerations. Thus, a displacement of
the nonlinearity from kinematic terms to dynamic ones weakens the
nonlinearity (Rylov, 1989).

The complex variable $\Psi $ is used in quantum mechanics, where it is known
as the wave function. The fact that the Schr\"odinger equation for the wave
function describes a irrotational flow of some ideal fluid is well known
(Madelung, 1926). For such a fluid the internal energy depends on $\rho $
and $\nabla \rho $. This internal energy has such a form $(E=b^{2}(\nabla
\rho )^{2}/2\rho ^{2}$, $b=-\hbar /m)$ that rhs of (8.3) vanishes, and the
equation (8.3) becomes linear.

In the case of incompressible fluid rhs of (8.3) becomes indefinite due to
additional constraint. The irrotational flow of the incompressible fluid is
described by the equations
$$
ib{\frac{\partial \Psi }{\partial t}}+{\frac{b^{2}}{2}}\nabla ^2\Psi = {%
\frac{p}{\rho _0}} \Psi , \eqno (8.4)
$$
$$
\bar{\Psi }\Psi =\rho _{0}=\hbox{const } \eqno (8.5)
$$
\noindent where the pressure $p$ is considered as some function of time and
position. The situation is the same as in the case of equations (6.8), where
the value of the pressure $p$ is determined only after the flow is
calculated.

The equation (8.4) looks as a linear equation. But in reality it is
nonlinear because of the nonlinear constraint (8.5). In the case of a
rotational flow some additional nonlinear terms appear. However if the flow
is slightly rotational, there is a hope that these additional nonlinear
terms, as well the nonlinear constraint (8.5) could be considered as
corrections to the irrotational flow described by linear dynamic equation.

Using idea of linearization of the irrotational flow, let us introduce $n$%
-component complex function $\Psi =\{\Psi _{\alpha }\}, \alpha =1,2,\ldots n$%
, defining it by the relations
$$
\Psi _{\alpha }=\sqrt{\rho }e^{i\varphi /b}w_{\alpha }(\bxi ), \qquad \bar{%
\Psi }{}_{\alpha }=\sqrt{\rho }e^{-i\varphi /b}\bar{w}{}_{\alpha } (\bxi %
),\qquad \bar{\Psi }{}\Psi \equiv \sum^{n}_{\alpha =1}\bar{\Psi }{}_{\alpha
}\Psi _{\alpha } \eqno (8.6)
$$
\noindent where the bar over a symbol means the complex conjugate, $%
w_{\alpha }(\bxi )$, $\alpha =1,2,\ldots n$ are complex functions of only
variables $\bxi$, satisfying the relations
$$
-{\frac{ib}{2}}\sum^{n}_{\alpha =1}(\bar{w}{}_{\alpha }{\frac{\partial
w_{\alpha } }{\partial \xi _{\beta }}}-{\frac{\partial \bar{w}{}_{\alpha }}{%
\partial \xi _{\beta }}}w_{\alpha })=g^{\beta }(\bxi ),\qquad \beta
=1,2,3,\qquad \sum^{n}_{\alpha =1}\bar{w} _{\alpha }w_{\alpha }=1 \eqno (8.7)
$$
\noindent $n$ is such a natural number that equations (8.7) admit a
solution. Here $b$ is some constant of dimensionality $[b]=[L^2T^{-1}]$
which is introduced to make the quantities $w_\alpha$ and $\phi =\varphi /b$
dimensionless. The number $n$ and the form of functions $w$ depend on the
form of functions $g^\alpha$. But if $w_\alpha$, $\alpha =1,2,\ldots n$ is a
solution of (8.7), the variables $\Psi _\alpha$ are some functions of $\rho$%
, $\varphi$, $\bxi$, but not of their derivatives, although equations (8.7),
determining the transformation (8.6), contain derivatives of functions $%
w_\alpha$ with respect to $\bxi$.

It is easy to verify that $\rho$ and ${\rho {\bf v}}$ with ${\bf v}$ defined
by (1.11) have the form
$$
\rho =\bar{\Psi }\Psi, \qquad {\bf j}=\rho {\bf v}=-{\frac{ib}{2}}(\bar{\Psi
}{}\nabla\Psi -\nabla\bar{\Psi }{}\Psi ) -\eta\nabla S \bar{\Psi }\Psi \eqno %
(8.8)
$$
Then the variational problem with the action (3.20) appears to be equivalent
to the variational problem with the action functional
$$
{\cal A}[\Psi ,\bar{\Psi }{} ,\eta ,S]=\int \{{\frac{ib}{2}} (\bar{\Psi }{}%
\partial _{0}\Psi -\partial _{0}\bar{\Psi }\cdot\Psi ) +\eta\partial _0S
\bar{\Psi }\Psi
$$
$$
-{\frac{1}{2\bar{\Psi }{}\Psi }}[{\frac{ib}{2}}(\bar{\Psi }\nabla \Psi
-\nabla \bar{\Psi }\cdot\Psi )+\eta\nabla S\cdot\bar{\Psi }\Psi ] ^{2} -E[x,%
\bar{\Psi }{}\Psi ,S]\bar{\Psi }\Psi \}d^{4}x \eqno (8.9)
$$

Note that the function $\Psi $ considered as a function of independent
variables $\{t,{\bf x}\}$ is very indefinite in the sense that the same
fluid flow may be described by different $\Psi $-functions. There are two
reasons for such an indefiniteness. First, the functions $w_\alpha (\bxi )$
are not determined uniquely by differential equations (8.7). Second, their
arguments $\bxi$ as functions of $x$ are determined only to within the
transformation (1.16). Description of a fluid in terms of the function $\Psi
$ is more indefinite, than the description in terms of the hydrodynamic
potentials $\bxi$. Information about initial and boundary conditions
containing in the functions ${\bf g}(\bxi )$ is lost at the description in
terms of the $\Psi $-function. The two-component $\Psi $-function can be
obtained directly from the Clebsch variables by means of a proper change of
variables (Rylov, 1989). 
However, one cannot be sure that any flow can be described by two component
wave function. According to (8.7) the classification of wave functions over
the minimal number of components is connected with the form of functions ${%
\b g}(\bxi )$ and, hence with the integration of hydrodynamic equations.

Let the function $\Psi$ have $n$ components. Regrouping components of the
function $\Psi$ in the action (8.9), one obtains the action in the form
$$
{\cal A}_{E}[\Psi ,\bar{\Psi }{},\eta ,S]=\int \{ {\frac{1}{2}}[\bar{\Psi }(
ib\partial _{0} +A_0)\Psi +(-ib\partial _{0}\bar{\Psi }+A_0\bar{\Psi }) \Psi
]-
$$
$$
-{\frac{1}{2}}(ib\nabla\bar{\Psi } -{\bf A}\bar{\Psi })(-ib\nabla\Psi -{\bf A%
}\Psi )+
$$
$$
+{\frac{b^2}{4}}\sum\limits^n_{\alpha ,\beta =1}\bar{Q}_{\alpha\beta ,\gamma
} Q_{\alpha\beta ,\gamma }\rho +{\frac{b^2}{8\rho }}{(\nabla\rho)}^2-\rho
E\} d^{4}x,\qquad \rho\equiv\bar{\Psi }\Psi \eqno (8.10)
$$
where
$$
A=\{A_0,{\bf A}\},\qquad A_0\equiv\eta\partial _0S,\qquad {\bf A}\equiv\eta
\nabla S, \eqno (8.11)
$$
$$
Q_{\alpha\beta ,\gamma}={\frac{1}{\bar{\Psi }\Psi }}\left|
\begin{array}{cc}
\Psi _{\alpha} & \Psi _{\beta } \\
\partial _{\gamma }\Psi _{\alpha } & \partial _{\gamma }\Psi _{\beta }
\end{array}
\right| , \qquad \alpha ,\beta =1,2,\ldots n\qquad \gamma =1,2,3 \eqno (8.12)
$$

Corresponding dynamic equations have the form
$$
{\frac{\delta {\cal A}}{\delta \bar \Psi {}_\alpha }}=(ib\partial
_0+A_0)\Psi _\alpha -{\frac 12}(ib\nabla +{\bf A})^2\Psi _\alpha -{\frac{b^2}%
4}\sum_{\mu ,\nu =1}^n\bar Q_{\mu \nu ,\gamma }Q_{\mu \nu ,\gamma }\Psi
_\alpha
$$
$$
+{\frac{b^2}2}\sum_{\nu =1}^nQ_{\alpha \nu ,\gamma }\partial _\gamma \bar
\Psi {}_\nu +{\frac{b^2}2}\sum_{\nu =1}^n\partial _\gamma (Q_{\alpha \nu
,\gamma }\bar \Psi {}_\nu )+{\frac \partial {\partial \rho }}[{\frac{b^2}{%
8\rho }}(\nabla \rho )^2-\rho E]\Psi _\alpha
$$
$$
-\nabla ({\frac{b^2}{4\rho }}\nabla \rho )\Psi _\alpha =0,\qquad \alpha
=1,2,\ldots n\eqno (8.13)
$$
$$
{\frac{\delta {\cal A}}{\delta S}}=\partial _i(j^i\eta )-{\frac{\partial
(\rho E)}{\partial S}}=0,\eqno (8.14)
$$
$$
{\frac{\delta {\cal A}}{\delta \eta }}=-\partial _i(j^iS)=0,\eqno (8.15)
$$
\noindent where $j=\{\rho ,{\bf j}\}=\{j^k\}$, $k=0,1,2,3$ is defined by
(8.8).

In the case of the irrotational flow, when $g^\alpha (\bxi )=\partial \Phi (%
\bxi )/\partial \xi _\alpha $ equations (8.7) have a solution for $n=1,$ and
the function $\Psi $ may have one component. Then all $Q_{\alpha \beta
,\gamma }\equiv 0,$ as it follows from (8.12).

The number $n$ of the $\Psi $-function components in the actions (8.9) and
(8.10) is arbitrary. A formal variation of the action with respect to $\Psi
_{\alpha }$ and $\bar{\Psi }{}_{\alpha },\quad \alpha =1,2,\ldots n$ leads
to $2n$ real dynamic equaitons, but not all of them are independent. There
are such combinations of variations $\delta \Psi _{\alpha }$, $\delta \bar{%
\Psi }{}_ {\alpha }$, $\alpha =1,2,\ldots n$ which do not change expressions
(8.8) and $\bar{\Psi }\partial _0\Psi -\partial _0\bar{\Psi }\cdot\Psi$.
Such combinations of variations $\delta \Psi _{\alpha }$, $\delta \bar{\Psi
}{}_{\alpha }$, $\alpha =1,2,\ldots n$ do not change the action (8.9), and
corresponding combinations of dynamic equations $\delta {\cal A}/ \delta
\Psi _{\alpha }=0$,\quad $\delta {\cal A}/\delta \bar{\Psi }_{\alpha }=0$
are identities that associates with a dependence between dynamic equations.

Thus, increasing the number $n$, one increases the number of dynamic
equations, but the number of independent dynamic equations remains the same.
In such a situation it is important to determine the minimal number $n_m$ of
the $\Psi $-function components, sufficient for a solution of equations
(8.7) with the given vector field $g^{\beta }(\bxi )$ in the space $V_{\bxi %
} $ of the labels $\bxi $.

Note that under the relabeling transformations (1.16), the quantity ${\bf g}(%
\bxi )$ transforms as a vector
$$
g^{\beta }(\bxi )\to\tilde{g}^{\beta }(\tilde{\bxi })={\frac{\partial \xi
_{\alpha } }{\partial \tilde{\xi }_{\beta }}}g^{\alpha }(\bxi ), \qquad
\beta =1,2,3 \eqno (8.16)
$$
\noindent It is necessary for the quantities (8.8) and the action (8.9) to
be invariant with respect to the transformation (1.16)

Let ${\cal G}$ be a set of all vector fields $g^{\beta }(\bxi )$ in $V_{\bxi %
}$, and ${\cal G}_{n}$ be a set of such vector fields $g^{\beta }(\bxi )$ in
$V_{\bxi }$ which can be represented in the form

$$
g^{\beta }(\bxi )=\sum^{n}_{k=1}\eta ^{2}_{k}(\bxi ){\partial \zeta _{k}(%
\bxi )/ \partial \xi _{\beta }},\qquad \eta _{1}\equiv 1 \eqno (8.17)
$$
\noindent where $n$ is a fixed natural number, and the functions $\eta _{k}$%
, $\zeta _{k}$, $k=1,2,\ldots n$ are scalars in $V_{\bxi }$. Under the
relabeling transformation (1.16) the functions (8.17) transform as follows
$$
\eta _k(\bxi )\rightarrow\tilde{\eta }_k(\tilde{\bxi })=\eta _k(\bxi ),
\qquad \zeta _k(\bxi )\rightarrow\tilde{\zeta }_k(\tilde{\bxi })=\zeta _k(%
\bxi ), \qquad k=1,2,\ldots n
$$
$$
g^{\beta }(\bxi )\rightarrow\tilde{g}^{\beta }(\tilde{\bxi })= {\frac{%
\partial \xi _{\alpha }}{\partial \tilde{\xi }_{\beta }}}g^{\alpha }(\bxi ) =%
{\frac{\partial \xi _{\alpha }}{\partial \tilde{\xi }_{\beta }}}
\sum^{n}_{k=1}\eta ^{2}_{k}(\bxi ){\frac{\partial \zeta _{k}(\bxi )}{%
\partial \xi _{\alpha }}}=\sum^{n}_{k=1}\tilde{\eta }^{2}_{k}(\tilde{\bxi })
{\frac{\partial \tilde{\zeta }_{k}(\tilde{\bxi })}{\partial \tilde{\xi }_
{\alpha }}} \eqno (8.18)
$$
\noindent In other words, a vector field $g^{\beta }(\bxi )$ of the form
(8.17) transforms into the vector field $\tilde{g}^{\beta }(\tilde{\bxi })$
of the same form (8.17), and the set ${\cal G}_{n}$ is invariant with
respect to the group (1.16) of the relabeling transformations.

It is easy to see that
$$
{\cal G}_{n-1}\subseteq {\cal G}_{n},\qquad {\cal G}_{0}=\emptyset,\qquad
n=1,2,\ldots \eqno (8.19)
$$
\noindent because the $n$th term of the sum (8.17) can be combined with the
first one, if $\zeta _{n}$ is a function of $\eta _{n}$. Let
$$
{\cal S}_{n}={\cal G}_{n}\backslash {\cal G}_{n-1},\qquad n=1,2,\ldots \eqno %
(8.20)
$$
\noindent Then
$$
{\cal G}=\bigcup\limits^{s=n_m}_{s=1}{\cal S}_{s},\qquad {\cal S}%
_{l}=\emptyset , \qquad l=n_m+1,n_m+2,\ldots \eqno (8.21)
$$
\noindent where $n_m$ is the number of nonempty invariant subsets of the set
${\cal G}$. Each subset ${\cal S}_{k}$ contains only such vector fields $%
g^{\beta }(\bxi )$ which associate with the $k$-component $\Psi $-function $%
\Psi =\{\Psi _{\alpha }\},\quad \alpha =1,2,\ldots k$, having the components
$$
\Psi _{1}=\left\{ {(1-\sum^{k}_{\alpha =2}\eta ^2_\alpha )\rho}%
\right\}^{1/2} \exp [i(\phi +\zeta _{1})],
$$
$$
\Psi _{\alpha }=\eta _{\alpha }\sqrt{\rho } \exp [i(\phi +\zeta _{\alpha
}+\zeta _{1})],\qquad \alpha =2,3,\ldots k \eqno (8.22)
$$
In particular, the set ${\cal S}_{1}$ associates with an irrotational flow,
described by a one-component $\Psi $-function determined by one scalar $%
\zeta _1$; and the set ${\cal S}_{2}$ associates with a rotational flow
described by a two-component $\Psi $-function, determined by three scalar
functions $\zeta _{1}$, $\eta _{2}$, $\zeta _{2}$ (Clebsch variables).

Thus, types of the perfect fluid flows can be labeled by invariants of the
relabeling group (1.16). This labeling is connected with the minimal number $%
n_m$ of the wave function components. In the quantum mechanics the minimal
number $n_m$ of the wave function components is connected with the spin $s$
of a particle described by this wave function. The relation has the form
$$
s=(n_m-1)/2 \eqno (8.23)
$$
We shall refer to the quantity $s$ defined by (8.23) as the kinematic spin
(k-spin) of the fluid flow. According to this definition the kinematic spin
of an irrotational flow is equal to 0. The kinematic spin of the rotational
flow $s\ge 1/2$. The k-spin $s=1/2$ is possible. Is a flow of higher k-spin
possible? This interesting question is yet open. It seems to be connected
with the complicated problem of knottedness of vortex lines (Moffat, 1969;
Bretherton, 1970).

As far as the field ${\bf g}(\bxi )$ is determined by the velocity field $%
{\bf v}_{{\rm in}}({\bf x})$, at the initial moment, when $\bxi ={\bf x}$
and ${\bf g}({\bf x})= {\bf v}_{{\rm in}}({\bf x})$, the k-spin is
determined by the initial velocity ${\bf v}_{{\rm in}}({\bf x})$. In
particular, if the initial velocity can be represented in the Clebsch form
(4.31) with $\eta _{{\rm in}}=0$, the k-spin of such a flow is equal to $1/2$%
.

Let us consider a description of the incompressible fluid in terms of the
wave function. For the incompressible fluid the action (8.9) has the form
$$
{\cal A}[\bar \Psi ,\Psi ,p]=\rho _0\int \negthinspace \int \{{\frac{ib}2}%
(\bar \Psi \partial _0\Psi -\partial _0\bar \Psi \cdot \Psi )-\frac 12{\bf %
v^2+}P(1-\bar \Psi \Psi )\}dtd{\bf x},\eqno (8.24)
$$
\noindent where $\rho =\rho _0=$const is the fluid density,
$$
{\bf v}=-\frac{ib}2(\bar\Psi \nabla \Psi -\nabla \bar \Psi \cdot \Psi ) %
\eqno (8.25)
$$
is the fluid velocity, and $P$ is the Lagrange multiplier which introduces
the constraint
$$
{\rho /\rho }_0=\bar \Psi \Psi =1. \eqno (8.26)
$$
The dynamic equations are obtained as a result of a variation with respect
to $P$, $\bar \Psi $, $\Psi $. Variation of (8.24) with respect to $P$ gives
(8.26), and
$$
\rho _0^{-1}{\frac{\delta {\cal A}}{\delta \bar \Psi }}=ib\partial _0\Psi +i{%
\frac b2[}{\bf v}\nabla \Psi +\nabla ({\bf v}\Psi )]-P\Psi =0\eqno (8.27)
$$
Let us convolute $\bar \Psi $ with (8.27) and take imaginary and real parts
of the obtained relation. In virtue of (8.26) one obtains respectively
$$
\nabla {\bf v}=-\frac{ib}2\nabla (\bar \Psi \nabla \Psi -\nabla \bar \Psi
\cdot \Psi )=0 \eqno (8.28)
$$
$$
P=ib(\bar \Psi \partial _0 \Psi -\partial _0 \bar \Psi\cdot \Psi )-{\bf v}^2 %
\eqno (8.29)
$$
Then the dynamic equation (8.27) takes the form
$$
ib(\partial _0\Psi +{\bf v}\nabla \Psi )-P\Psi =0\eqno (8.30)
$$
where ${\bf v}$ is defined by (8.25). Let us use the expressions (8.6),
(8.7) of the wave function via functions ${\bf g}(\bxi )$. Using $\rho =1$,
one derives
$$
P=-\partial _0\varphi -g^\alpha (\bxi )\partial _0\xi _\alpha -{\bf v}^2 %
\eqno (8.31)
$$
Comparing (8.31) with (6.12), one derives the relation between $P$ and the
pressure $p$%
$$
\frac p{\rho _0}=P+{\bf v}^2/2 \eqno (8.32)
$$

\section{Perturbation Theory for the Flow of $k$-spin 1/2}

Let us consider the flow of k-spin $1/2$, when the wave function has the form
$$
\Psi =\left(
\begin{array}{c}
\beta  \\
\gamma
\end{array}\right),
\qquad \bar \Psi =\left( \bar\beta
,\bar\gamma \right) \eqno (9.1)
$$
If the wave function is one-component, k-spin is equal to 0 and the flow is
irrotational. If $\gamma \ll \beta $, the flow is slightly rotational. The
expanded form of (8.30)
$$
ib(\partial _0\beta +{\bf v}\nabla \beta )-P\beta =0,\qquad ib(\partial
_0\gamma +{\bf v}\nabla \gamma )-P\gamma =0\eqno (9.2)
$$
$$
{\bf v}=-\frac{ib}2(\bar{\beta }\nabla \beta +\bar{\gamma }\nabla \gamma
-\nabla \bar{\beta }\cdot \beta -\nabla \bar{\gamma }\cdot \gamma )\eqno %
(9.3)
$$

Let the expansion of dynamic variables $\beta ,$ $\gamma ,$ {\bf v}$,$ $P$
have the form
$$
\beta =\beta _0+\epsilon \beta _1+\epsilon ^2\beta _2+O(\epsilon ^3), \qquad
\gamma =\epsilon\gamma _1+\epsilon ^2\gamma _2+O(\epsilon ^3),
$$
$$
P=P_0+\epsilon P_1+\epsilon ^2P_2+O(\epsilon ^3),\qquad {\bf v}={\bf v}%
_0+\epsilon {\bf v}_1+\epsilon ^2{\bf v}_2+O(\epsilon ^3),\qquad \epsilon
\ll 1\eqno (9.4)
$$
\noindent where $\epsilon $ is a formal small parameter. The expansion is
used to within $\epsilon ^3$, because it appears that the expansion of
variables $\beta ,$ {\bf v}$,$ $P$ contains only even powers of $\epsilon $,
whereas the expansion of $\gamma $ contains only odd powers of $\epsilon $.

Let us substitute expansions (9.4) into dynamic equations (8.26), (8.28),
(8.30) and equate to zero coefficients before different powers of $\epsilon$%
. In the zeroth order approximation one derives
$$
\bar \beta _0\beta _0=1,\qquad \beta _0=e^{i\phi _0},\qquad {\bf v}%
_0=b\nabla \phi _0\eqno (9.5)
$$
where $\phi _0$ is a real variable. Then the equations (8.28) and (8.30)
give respectively
$$
\nabla ^2\phi _0=0,\qquad P_0=-b\partial _0\phi _0-b^2(\nabla \phi
_0)^2,\qquad \eqno (9.6)
$$

\noindent Thus in the zeroth approximation we have a problem of an
irrotational incompressible flow. This problem is supposed to be solved.

In the first order approximation we have
$$
\bar \beta _0\beta _1+\bar \beta _1\beta _0=0\qquad \beta _1=i\phi _1\beta _0%
\eqno (9.7)
$$
\noindent where $\phi _1$ is a real quantity. Substituting (9.7) into
(8.28), (9.2), one derives

$$
\nabla ^2\phi _1=0,\qquad P_1=-b\partial _0\phi _1-b^2\nabla \phi _0\nabla
\phi _1,\eqno (9.8)
$$
\noindent As far as the dynamic equation for $\phi _1$ has the same form as
for $\phi _0$, one can replace $\phi _0+\varphi \phi _1$, by one variable $%
\phi _0$. It means that formally one may set without a loss of generality
$$
\phi _1=0,\qquad P_1=0,\qquad \beta _1=0\eqno (9.9)
$$
\noindent The second equation (9.2) reduces to the form
$$
i(\partial _0\gamma _1+b\nabla \phi _0\nabla \gamma _1)+[\partial _0\phi
_0+b(\nabla \phi _0)^2]\gamma _1=0\eqno (9.10)
$$
It is easy to see that (9.10) reduces to the equation of the type of (7.10).
The general solution of (9.10) has the form
$$
\gamma _1=R_1(X-u_\infty t,\eta ,\zeta )\exp [{i\phi _0+i\vartheta
_1(X-u_\infty t,\eta ,\zeta )]}\eqno (9.11)
$$
where $R_1$, $\vartheta _1$ are arbitrary functions of arguments $X-u_\infty
t,\eta ,\zeta $ defined by relations (7.4)-(7.6) with $\varphi _0=b\phi _0$.

Taking into account (9.9), one has in the second order approximation
$$
\bar \beta _0\beta _2+\bar \beta _2\beta _0+\bar \gamma _1\gamma _1=0\eqno %
(9.12)
$$
This equation has the general solution
$$
\beta _2=(-{\frac 12}\bar \gamma _1\gamma _1+i\phi _2)\beta _0\eqno (9.13)
$$
where $\phi _2$ is a real variable. Equation for $\gamma _2$ has the form
$$
i(\partial _0\gamma _2+b\nabla \phi _0\nabla \gamma _2)+[\partial _0\phi
_0+b(\nabla \phi _0)^2]\gamma _2=0\eqno (9.14)
$$
which coincides with the equation (9.10) for $\gamma _1$. It means the $%
\epsilon\gamma _2$ may be included in $\gamma _1$ and without a loss of
generality one may set $\gamma _2=0$.

The equation (8.25) and (8.28) are written in the form
$$
{\bf v}_2=b\nabla \phi _2+bR_1^2\nabla \vartheta _1,\qquad \nabla ^2\phi
_2=-\nabla (R_1^2\nabla \vartheta _1)\eqno (9.15)
$$
Then the first equation (9.2) gives

$$
P_2=-b(\partial _0\phi _2+{\bf v}_0\nabla \phi _2+{\bf v}_2\nabla \phi _0) %
\eqno (9.16)
$$

From (9.5), (9.15) one derives
$$
{\bf v}=\nabla (\varphi _0+\epsilon ^2\varphi _2)+\epsilon ^2bR_1^2\nabla
\vartheta _1+O(\epsilon ^3),\qquad \varphi _0=b\phi _0,\qquad \varphi
_2=b\phi _2\eqno (9.17)
$$
Using (8.32), (9.6), (9.8), (9.9), (9.16), (9.17), one derives after
calculations
$$
\frac p{\rho _0}=-\partial _0(\varphi _0+\epsilon ^2\varphi _2)-\frac
12(\nabla \varphi _0)^2-\epsilon ^2\nabla \varphi _0\nabla \varphi
_2+O(\epsilon ^3),\eqno (9.18)
$$
Results (9.17), (9.18) agree with (7.23), (7.24), provided one sets $%
\epsilon ^2=\varepsilon $. The expression (7.23) contains the combination $%
A_\alpha \nabla B_\alpha $, $\alpha =1,2,3$, whereas the expression (9.17)
contains only two arbitrary functions $R_1$, $\vartheta _1$. The fact is
that that (7.23) describes arbitrary slightly rotational flow, whereas the
relation (9.17) describes slightly rotational flow of k-spin $1/2$. For any
flow of k-spin $s\leq 1/2,$ the six functions $A_\alpha $, $B_\alpha $, $%
\alpha =1,2,3$ reduce to two functions (for instance, $A_1$, $B_1$). But the
question about maximal value of k-spin is open now. On one hand, one can
prove (see, for instance, appendix to the paper by Eckart, 1960) that any
vector field ${\bf g}({\bf x})$ in the three-dimensional space can be
represented in the Clebsch form
$$
{\bf g}=\nabla \zeta _1+\eta _2\nabla \zeta _2 \eqno (9.19)
$$
On the other hand there are examples of vector fields which cannot be
represented in the form (9.19) inside the whole space (Moffat, 1969), if $%
\zeta _1$, $\eta _2$, $\zeta _2$ are considered as single-valued functions
of coordinates.

\section{Two-dimensional irrotational flow}

Let us consider a flow around a circular cylinder of a radius $a$. The
cylinder axis is supposed to coincide with the axis of $z$. The flow is
directed along the axis of $x$. The basic irrotational flow is described by
the potential $\varphi _0$, and by the stream function $\psi $ which are
connected with the velocity components ${\bf v}_0=\{v_x,v_y,v_z\}$ by means
of relations
$$
v_x=\partial _x\varphi _0=\partial _y\psi ,\qquad v_y=\partial _y\varphi
_0=-\partial _x\psi ,\qquad v_z=0\eqno (10.1)
$$
We are going to use variables $\varphi ,\psi $ as coordinates, and introduce
another variables $\xi ,\eta ,\zeta $ defined by (7.4). In the given case
they are determined by relations
$$
\xi =u_\infty ^{-1}\varphi _0=bu_\infty ^{-1}\phi _0,\qquad \eta =bu_\infty
^{-1}\psi ,\qquad \zeta =z\eqno (10.2)
$$
\noindent where $u_\infty $ is the velocity of the fluid at infinity. In
absence of a circulation one derives for the circular cylinder (see, for
instance Lamb, 1932)
$$
\xi =(1+{\frac{a^2}{r^2}})r\cos \vartheta ,\qquad \eta =(1-{\frac{a^2}{r^2}}%
)r\sin \vartheta ,\eqno (10.3)
$$
$$
r^2=x^2+y^2,\qquad \vartheta =\arctan {\frac yx}
$$
$$
\partial _x\xi =\partial _y\eta ,\qquad \partial _y\xi =-\partial _x\eta
\qquad \nabla ^2\xi =0,\qquad \nabla ^2\eta =0,\eqno (10.4)
$$
\noindent In virtue of (10.4)
$$
\nabla \xi \cdot \nabla \eta =0,\qquad (\nabla \xi )^2=V_0^2=(\nabla \eta
)^2={\frac{\partial (\xi ,\eta )}{\partial (x,y)}}\equiv D_0\eqno (10.5)
$$
\noindent where $V_o=\left| \nabla \xi \right| /u_\infty $ is a
dimensionless velocity. For the case of the circular cylinder the
determinant
$$
V_0^2=D_0=1+{\frac{a^4}{r^4}}-{\frac{2a^2}{r^2}}\cos (2\vartheta )\eqno %
(10.6)
$$
\noindent tends to 1, if $r\rightarrow \infty $.

Let us represent the variable $X$ in the form
$$
X=\xi +\Delta ,\qquad \Delta =\Delta (\xi ,\eta )=\int\limits_0^\xi
[D_0^{-1}(\xi ^{\prime },\eta )-1]d\xi ^{\prime }\eqno (10.7)
$$
Then according to (10.6)
$$
D_0^{-1}(\xi ,\eta )-1=-\frac{a^2/r^2-2\cos (2\vartheta )}{%
r^2/a^2+a^2/r^2-2\cos (2\vartheta )}\eqno (10.8)
$$
At $\xi \to \pm \infty $
$$
\Delta (\xi ,\eta )=\Delta _{\pm }(\eta )+O(\xi ^{-1}),\qquad \xi \to \pm
\infty \eqno (10.9)
$$

Let us try to find a slightly rotational stationary flow. If the flow is
stationary, the functions $A_\alpha $ in (7.23) do not depend on argument $%
X-u_\infty t$. $B_\alpha $ are linear with respect to $X-u_\infty t$ and do
not depend on $\eta ,\zeta $. Let the flow do not depend on $\zeta =z$. Then
$$
A_1=f(\eta ),\qquad f(\eta )\rightarrow 0,\qquad \eta \to \infty \qquad
A_2=A_3=0\eqno (10.10)
$$
$$
B_1=X-u_\infty t,\qquad B_2=B_3=0\eqno (10.11)
$$
where $X$ is determined by (7.6). Then according to (7.23), (10.2), one
derives
$$
{\bf v}=u_\infty [1+\varepsilon f(\eta )]\nabla \xi +\varepsilon f(\eta
)\nabla \Delta (\xi ,\eta )+\varepsilon \nabla \varphi _1\eqno (10.12)
$$
The potential $\varphi $ is determined by the relation (7.22) which takes
the form
$$
\varphi _1(\xi ,\eta )=-\frac{u_\infty }{4\pi }\int \nabla ^{\prime }G({\bf x%
},{\bf x^{\prime }})f(\eta ^{\prime })\nabla ^{\prime }\Delta (\xi ^{\prime
},\eta ^{\prime })d{\bf x}^{\prime }\eqno (10.13)
$$
where $\xi ,\eta $ are functions (10.3) of ${\bf x}=\{x,y,z\}$, and $\xi
^{\prime },\eta ^{\prime }$ are functions (10.3) of ${\bf x}^{\prime
}=\{x^{\prime },y^{\prime },z^{\prime }\}$. Introducing designation
$$
\tilde \nabla =\{{\frac \partial {\partial \xi }},{\frac \partial {\partial
\eta }},D_0^{-1/2}{\frac \partial {\partial \zeta }}\}\eqno (10.14)
$$
and using (10.4), (10.5), the relation (10.13) reduces to
$$
\varphi _1(\xi ,\eta )=-\frac{u_\infty }{4\pi }\int \tilde \nabla ^{\prime
}G(\xi ,\eta ;\xi ^{\prime },\eta ^{\prime })f(\eta ^{\prime })\tilde \nabla
^{\prime }\Delta (\xi ^{\prime },\eta ^{\prime })d\xi ^{\prime }d\eta
^{\prime }\eqno (10.15)
$$
where the Green function $G(\xi ,\eta ;\xi ^{\prime },\eta ^{\prime })$ has
the form
$$
G(\xi ,\eta ;\xi ^{\prime },\eta ^{\prime })=-2\ln \sqrt{(\xi -\xi ^{\prime
})^2+(\eta -\eta ^{\prime })^2}-2\ln \sqrt{(\xi -\xi ^{\prime })^2+(\eta
+\eta ^{\prime })^2}\eqno (10.16)
$$
Such a reduction of the Green function is possible only in the case, when $%
\Delta (\xi ,\eta )$ does not depend on $\zeta =z$. Here $\xi $, $\eta $ are
functions of $(x,y)$ defined by (10.3), and $\xi ^{\prime }$, $\eta ^{\prime
}$ are the same functions of $(x^{\prime },y^{\prime })$. Calculation of $D_0
$ and of $\partial\Delta /\partial \xi $ in terms of variables $\xi ,\eta
,\zeta $ gives
$$
D_0(\xi ,\eta )=\frac{4\sqrt{1+\alpha _1^2-2\alpha _2}}{\alpha _1+\sqrt{%
1+\alpha _1^2-2\alpha _2}+\sqrt{2(\alpha _1^2+\alpha _1\sqrt{1+\alpha
_1^2-2\alpha _2}-\alpha _2)}}\eqno (10.17)
$$
$$
\frac{\partial \Delta }{\partial \xi }=\frac{\alpha _1+\sqrt{2(\alpha
_1^2+\alpha _1\sqrt{1+\alpha _1^2-2\alpha _2}-\alpha _2)}}{4\sqrt{1+\alpha
_1^2-2\alpha _2}}-\frac34 \eqno (10.18)
$$
$$
\alpha _1=(\xi ^2+\eta ^2)/4a^2,\qquad \alpha _2=(\xi ^2-\eta ^2)/4a^2\eqno %
(10.19)
$$

Using the relations (7.24), (10.2), (10.5), (10.12), (10.15) and the second
relation (7.20) written in the form
$$
\frac{\partial G}{\partial \eta }\left. (\xi ,\eta ;\xi ^{\prime },\eta
^{\prime })\right| _{\eta =0}=0\eqno (10.20)
$$
one obtains for the pressure $p$%
$$
\frac p{\rho _0}=-\frac 12\left( \nabla \varphi _0\right) ^2-\varepsilon
\nabla \varphi _0\nabla \varphi _1+O(\varepsilon ^2)=
$$
$$
u_\infty ^2D_0(\xi ,\eta )\left[ -\frac 12+\frac \varepsilon {4\pi }\frac
\partial {\partial \xi }\int (\frac{\partial G}{\partial \xi ^{\prime }}%
\frac{\partial \Delta ^{\prime }}{\partial \xi ^{\prime }}+\frac{\partial G}{%
\partial \eta ^{\prime }}\frac{\partial \Delta ^{\prime }}{\partial \eta
^{\prime }})f(\eta ^{\prime })d\xi ^{\prime }d\eta ^{\prime }\right] \eqno %
(10.21)
$$
$$
\Delta ^{\prime }=\Delta (\xi ^{\prime },\eta ^{\prime }),\qquad G=G(\xi
,\eta ;\xi ^{\prime },\eta ^{\prime })\eqno (10.22)
$$

Now let us calculate the force {\bf F} acting on the cylinder. Components of
the external normal ${\bf n}$ are described by relations
$$
{\bf n}=\{n_x,n_y\}=\{\cos \vartheta ,\sin \vartheta \}=\{\xi /2a,\sqrt{%
1-\xi ^2/4a^2}\}\eqno (10.23)
$$
Components of the force ${\bf F}$ have the form
$$
F_x=-\int\limits_{-\pi }^\pi p_0(\vartheta )n_xad\vartheta =\oint \frac{%
p(\xi )d\xi }{\sqrt{1-\xi ^2/4a^2}},\qquad F_y=-\int\limits_{-\pi }^\pi
p_0(\vartheta )n_yad\vartheta =\oint p(\xi )d\xi \eqno (10.24)
$$
where the contour integral means the integral in the complex plane of $\xi
+i\eta $, taken around the cut $\eta =0$ between the points $a$ and $-a$. $%
p(\xi )=p_0(\theta )$ is the pressure on the surface of the cylinder.

Using expression (10.21) for the pressure $p$, one derives
$$
F_x=-\rho _0u_\infty ^2\int\limits_{-\pi }^\pi \cos \vartheta ad\vartheta
\left\{ D_0(\xi ,0)\left[ -\frac 12+\frac \varepsilon {4\pi }M(\xi )\right]
\right\} +O(\varepsilon ^2),\eqno (10.25)
$$
$$
F_y=-\rho _0u_\infty ^2\int\limits_{-\pi }^\pi \sin \vartheta ad\vartheta
\left\{ D_0(\xi ,0)\left[ -\frac 12+\frac \varepsilon {4\pi }M(\xi )\right]
\right\} +O(\varepsilon ^2),\eqno (10.26)
$$
where
$$
M(\xi )=\frac \partial {\partial \xi }\int\limits_{-\infty }^\infty
\int\limits_{-\infty }^\infty \left( \frac{\partial G}{\partial \xi ^{\prime
}}\frac{\partial \Delta ^{\prime }}{\partial \xi ^{\prime }}+\frac{\partial G%
}{\partial \eta ^{\prime }}\frac{\partial \Delta ^{\prime }}{\partial \eta
^{\prime }}\right) f(\eta ^{\prime })d\xi ^{\prime }d\eta ^{\prime },\eqno %
(10.27)
$$
$\xi =2a\cos \vartheta $ and $D_0(\xi ,0)=4\sin ^2\vartheta $. The first
term in (10.25), (10.26) vanishes and
$$
F_x=-\frac{\varepsilon \rho _0u_\infty ^2}{4\pi }\int\limits_{-\infty
}^\infty \int\limits_{-\infty }^\infty \left( \frac{\partial I_x}{\partial
\xi ^{\prime }}\frac{\partial \Delta ^{\prime }}{\partial \xi ^{\prime }}+%
\frac{\partial I_x}{\partial \eta ^{\prime }}\frac{\partial \Delta ^{\prime }%
}{\partial \eta ^{\prime }}\right) f(\eta ^{\prime })d\xi ^{\prime }d\eta
^{\prime }+O(\varepsilon ^2),\eqno (10.28)
$$
$$
F_y=-\frac{\varepsilon \rho _0u_\infty ^2}{4\pi }\int\limits_{-\infty
}^\infty \int\limits_{-\infty }^\infty \left( \frac{\partial I_y}{\partial
\xi ^{\prime }}\frac{\partial \Delta ^{\prime }}{\partial \xi ^{\prime }}+%
\frac{\partial I_y}{\partial \eta ^{\prime }}\frac{\partial \Delta ^{\prime }%
}{\partial \eta ^{\prime }}\right) f(\eta ^{\prime })d\xi ^{\prime }d\eta
^{\prime }+O(\varepsilon ^2),\eqno (10.29)
$$
where
$$
I_x(\xi ^{\prime },\eta ^{\prime })=\int\limits_{-\pi }^\pi D_0(\xi ,0)\frac{%
\partial G(\xi ,0;\xi ^{\prime },\eta ^{\prime })}{\partial \xi }a\cos
\vartheta d\vartheta \eqno (10.30)
$$
$$
I_y(\xi ^{\prime },\eta ^{\prime })=\int\limits_{-\pi }^\pi D_0(\xi ,0)\frac{%
\partial G(\xi ,0;\xi ^{\prime },\eta ^{\prime })}{\partial \xi }a\sin
\vartheta d\vartheta \eqno (10.31)
$$
$$
G(\xi ,0;\xi ^{\prime },\eta ^{\prime })=-2\ln \left[ (\xi -\xi ^{\prime
})^2+\eta ^{\prime 2}\right] =-2\ln \left[ (2a\cos \vartheta -\xi ^{\prime
})^2+\eta ^{\prime 2}\right] \eqno (10.32)
$$
It is easy to verify that $I_y(\xi ^{\prime },\eta ^{\prime })=0.$ Then $%
F_y=0$.

$I_x(\xi ,\eta )$ is a $\sin $gle-valued function of $\arg $uments $\xi ^2$
and $\eta ^2$ (see details in Appendix). As it follows from (10.6), (10.17),
(10.18) $\partial \Delta /\partial \xi $ is a single-valued function of
arguments $\xi ^2$ and $\eta ^2$, and $\Delta $ has the form
$$
\Delta (\xi ,\eta )=\xi B(\xi ^2,\eta ^2)\eqno (10.33)
$$
where $B$ is a single-valued function of arguments $\xi ^2$ and $\eta ^2.$
Then
$$
\frac{\partial I_x}{\partial \xi }\frac{\partial \Delta }{\partial \xi }=\xi
C_1(\xi ^2,\eta ^2),\qquad \frac{\partial I_x}{\partial \eta }\frac{\partial
\Delta }{\partial \eta }=\xi C_2(\xi ^2,\eta ^2)\eqno (10.34)
$$
where $C_1$ and $C_2$ are single-valued functions of $\arg $uments $\xi ^2$
and $\eta ^2$. It follows from (10.28) and (10.34) that $F_x$ also vanish in
the first order approximation $F_x=O(\varepsilon ^2)$. In other words, the
D'Alamberian paradox takes place also in the case of the slightly rotational
stationary flow.

\section{Concluding remarks}

Basing on the Jacobian technique and on the invariance with respect to the
relabeling group, one succeeded to integrate dynamic equtions for the ideal
fluid. Three indefinite functions ${\bf g}(\bxi )$ of three arguments arise
as a result of this integration. These functions can be expressed vias
initial and boundary conditions, and all essential information on the fluid
flow appears to be concentrated in the integrated system of hydrodynamic
equations.

Although the integrated system and the curtailed Euler system contain the
same number of equations, they differ in the relation that the integrtated
system takes into account information contained in the Lin constraints,
whereas the Euler system ignores it. This difference is of no importance for
irrotational flows, when the fluid properties described by the Kelvin's
theorem (and Lin constraints) are inessential, because they are fulfilled
automatically. But this difference is essential in the case of irrotational
flows.

One should expect that in the case of a compressible fluid the hydrodynamic
potentials and the wave function will describe effectively an interaction
between the acoustic waves and the slight fluid vorticity.

Integration of dynamic equations generates two ways of descriptions: a
description in terms of hydrodynamic potential (DTHP) and a description in
terms of wave function (DTWF). Both DTHP and DTWF take into account Lin
constraints and appear to be effective in the case of slightly rotational
flows. There is a hope they will appear to be effective also in the case of
strongly rotational flows. Maybe, an application of DTHP and DTWF will allow
to formulate the turbulence problem in a proper way. At any rate DTHP and
DTWF seem to be more effective for rotational flows, than the conventional
description in terms of velocity, because they take into account information
contained in Lin constraints (and Kelvin's theorem). DTHP can take into
account even initial and boundary conditions.

The integration of hydrodynamic equations and appearance of the field ${\bf g%
}$ activates the relabeling group. Invariant subsets of this group can be
used for a classification of the fluid flows. The field ${\bf g}$ appears to
be a tool for introducing such attributes of the ideal fluid as the wave
function and the spin. These concepts are new for conventional
hydrodynamics, although they are well known in quantum mechanics. In some
cases it may be useful in the sense that quantum mechanical methods can be
used in hydrodynamics and vice versa. \bigskip
\bigskip

The author is indebted to Prof. V.A.Gorodtsov for fruitful discussions which
were very useful for writing this paper. \newpage

\section{Appendix}

Let us calculate the expression (10.30)
$$
I_x(\xi ,\eta )=\int\limits_{-\pi }^\pi D_0(2a\cos \vartheta ,0)\left. \frac{%
\partial G(\xi ^{\prime },0;\xi ,\eta )}{\partial \xi ^{\prime }}\right|
_{\xi ^\prime =2a\cos \vartheta }a\cos \vartheta d\vartheta \eqno (A.1)
$$
where
$$
G(2a\cos \vartheta ,0;\xi ,\eta )=-2\ln \left[ (2a\cos \vartheta -\xi
)^2+\eta ^2\right] \eqno (A.2)
$$
$$
\frac{\partial G}{\partial \xi ^{\prime }}=-\frac 1{2a\sin \vartheta }\frac{%
\partial G}{\partial \vartheta }=\frac{4(2a\cos \vartheta -\xi )}{(2a\cos
\vartheta -\xi )^2+\eta ^2}\eqno (A.3)
$$
Let us introduce designations
$$
\mu =\mu _1+i\mu _2,\qquad \stackrel{}{\bar{\mu }}=\bar{\mu }_1+i\bar{\mu }%
_2,\qquad \mu _1=\frac \xi {2a},\qquad \mu _2=\frac \eta {2a}\eqno (A.4)
$$
Then
$$
D_0(2a\cos \vartheta ,0)=4\sin ^2\vartheta \eqno (A.5)
$$
$$
I_x(\xi ,\eta )=8\int\limits_{-\pi }^\pi \frac{\cos \vartheta -\mu _1}{(\cos
\vartheta -\mu _1)^2+\mu _2^2}(1-\cos ^2\vartheta )\cos \vartheta d\vartheta
=
$$
$$
4\int\limits_{-\pi }^\pi (1-\cos ^2\vartheta )\cos \vartheta \left( \frac
1{\cos \vartheta -\mu _1+i\mu _2}+\frac 1{\cos \vartheta -\mu _1-i\mu
_2}\right) d\vartheta \eqno (A.6)
$$
This integral reduces to the form
$$
I_x=-16A\pi -4A\int\limits_{-\pi }^\pi \left( \frac \mu {\cos \vartheta -\mu
}+\frac{\bar{\mu }}{\cos \vartheta -\bar{\mu }}\right) d\vartheta \eqno %
(A.7)
$$
where
$$
A=(\mu ^2+\bar{\mu }^2+\mu \bar{\mu }-1)=3\mu _1^2-\mu _2^2-1\eqno (A.8)
$$
Integration of (A.7) leads to
$$
I_x=-16A\pi -4A\left[ \frac \mu {\sqrt{1-\mu ^2}}\left. \ln \frac{\tan \frac
\vartheta 2+\frac{1-\mu }{\sqrt{1-\mu ^2}}}{\tan \frac \vartheta 2-\frac{%
1-\mu }{\sqrt{1-\mu ^2}}}\right| _{-\pi }^\pi +(c.c.)\right] \eqno (A.9)
$$
where ''(c.c)'' means the complex conjugate quantity. Let us introduce the
complex quantity
$$
\nu =\nu _1+i\nu _2=\frac \mu {\sqrt{1-\mu ^2}}=\sqrt{\frac{1-\mu }{1+\mu }}%
\eqno (A.10)
$$
If $\vartheta $ changes from$-\pi $ to $\pi $, the changes of arguments of
complex quantities $\tan \frac \vartheta 2+\nu $ and $\tan \frac \vartheta
2-\nu $ are respectively
$$
\Delta \arg (\tan \frac \vartheta 2+\nu )=-\pi {\rm sgn(Im}\nu ),\qquad
\Delta \arg (\tan \frac \vartheta 2-\nu )=\pi {\rm sgn(Im}\nu )\eqno (A.11)
$$
Then
$$
I_x=-16A\pi +8A\pi i\left[ \frac \mu {\sqrt{1-\mu ^2}}-(c.c.)\right] {\rm sgn%
} \nu _2\eqno (A.12)
$$
Let
$$
\mu =\mu _1+i\mu _2=\cos (\lambda _1+i\lambda _2)=\cos \lambda \eqno (A.13)
$$
where $\mu _1$, $\mu _2$, $\lambda _1$, $\lambda _2$ are real quantities.
Then according to (A.10)
$$
\nu =\nu _1+i\nu _2=\sqrt{\frac{1-\cos \lambda }{1+\cos \lambda }}=\tan
\frac \lambda 2=\frac{\sin \lambda _1}{\cos \lambda _1+\cosh \lambda _2}+i%
\frac{\sinh \lambda _2}{\cos \lambda _1+\cosh \lambda _2}\eqno (A.14)
$$
It follows from (A.13)
$$
\mu _1=\cos \lambda _1\cosh \lambda _2,\qquad \mu _2=\sin \lambda _1\sinh
\lambda _2\eqno (A.15)
$$
Or
$$
\sin ^2\lambda _1=\left| q\right| ^{1/2}-\frac{\mu _1^2+\mu _2^2-1}2,\qquad
\cos ^2\lambda _1=\frac{\mu _1^2+\mu _2^2+1}2-\left| q\right| ^{1/2}\eqno %
(A.16)
$$
$$
\sinh ^2\lambda _2=\frac{\mu _1^2+\mu _2^2-1}2+\left| q\right| ^{1/2},\qquad
\cosh ^2\lambda _2=\frac{\mu _1^2+\mu _2^2+1}2+\left| q\right| ^{1/2}\eqno %
(A.17)
$$
where
$$
q\equiv \left( \frac{\mu _1^2+\mu _2^2+1}2\right) ^2-\mu _1^2\equiv \left(
\frac{\mu _1^2+\mu _2^2-1}2\right) ^2+\mu _2^2\eqno (A.18)
$$
and the square root of the module is supposed to be non-negative.

All functions (A.16), (A.17) are non-negative. They vanish only at the
folowing conditions: sin$^2\lambda _1=0$ at $\mu _2=0\wedge \mid \mu _1\mid
\geq 1$, \quad cos$^2\lambda _1=0$ at $\mu _1=1$,\quad sinh$^2\lambda _2=0$
at $\mu _2=0\wedge \mid \mu _1\mid \leq 1$. It means that sinh$\lambda _2$
is a single-valued function on the complex plane $\mu $ with the cut $\mu
_2=0\wedge \mid \mu _1\mid \leq 1$, and the sign of $\nu _2$%
$$
{\rm sgn}\nu _2={\rm sgn}\frac{\sinh \lambda _2}{\cos \lambda _1+\cosh
\lambda _2}={\rm sgn}({\rm sinh}\lambda _2)\eqno (A.19)
$$
is the same on the whole complex plane $\mu $, because $\mid \cos \lambda
_1\mid \leq \cosh \lambda _2$.\quad $\cos \lambda _1$ may coincide with $%
-\cosh \lambda _2$ only if sin$^2\lambda _1=0$, and hence $\mu _2=0\wedge
\mid \mu _1\mid \geq 1$. In this case $\cosh ^2\lambda _2=(\mu
_1^2+1)/2+\mid (\mu _1^2-1)/2\mid >1$ for $\mu _1^2>1$.\quad $\cosh
^2\lambda _2=1$ only if $\mu _2=0\wedge \mu _1=\pm 1$, but these points $\mu
=1$ and $\mu =-1$ lie on the cut and are supposed to be inaccessible. Thus $%
\nu _2$ has the same sign on the whole complex plane $\mu $ with the cut $%
\mu _2=0\wedge \mid \mu _1\mid \leq 1$. Expression in the brackets of (A.12)
can be represented as follows
$$
\frac \mu {\sqrt{1-\mu ^2}}-(c.c.)=\frac{\cos \lambda _1}{\sin \lambda _1}-%
\frac{\cos \bar{\lambda }_1}{\sin \bar{\lambda }_1}=\tan (\frac \pi
2-\lambda )-\tan (\frac \pi 2-\bar{\lambda })=
$$
$$
-i\frac{2\sinh \lambda _2\cosh \lambda _2}{\cosh ^2\lambda _2+\sinh
^2\lambda _2-\cos ^2\lambda _1+\sin ^2\lambda _1}\eqno (A.20)
$$
Using (A.16), (A.17), one obtains
$$
\cosh ^2\lambda _2+\sinh ^2\lambda _2-\cos ^2\lambda _1+\sin ^2\lambda
_1=4\left[ \left( \frac{\mu _1^2+\mu _2^2-1}2\right) ^2+\mu _2^2\right]
^{1/2}\eqno (A.21)
$$

$$
\sinh ^2\lambda _2\cosh ^2\lambda _2=\left( \frac{\mu _1^2+\mu _2^2-1}%
2\right) ^2+\mu _2^2+\left( \frac{\mu _1^2+\mu _2^2}2\right) ^2
$$
$$
-\frac14+(\mu _1^2+\mu _2^2)\sqrt{\left( \frac{\mu _1^2+\mu _2^2-1}2 \right)
^2+\mu _2^2} \eqno (A.22)
$$
(A.18) is positive everywhere on the plane $\mu$ with the cut $\mu
_2=0\wedge \mid \mu _1\mid \leq 1$ and
$$
I_x=-16A\pi +4A\pi \frac{\sqrt{2\left( \frac{\mu _1^2+\mu _2^2}2\right)
^2-\mu _1^2+(\mu _1^2+\mu _2^2)\sqrt{\left( \frac{\mu _1^2+\mu _2^2-1}%
2\right) ^2+\mu _2^2}}}{\left| \left( \frac{\mu _1^2+\mu _2^2-1}2\right)
^2+\mu _2^2\right| }{\rm sgn}({\rm sinh}\lambda _2)\eqno (A.23)
$$
is a single-valued function of arguments $\mu _1^2=\xi ^2/4a^2$ and $\mu
_2^2=\eta ^2/4a^2.$

\end{document}